
\input harvmac
\noblackbox
\def \tes {\textstyle}
\def \a {\alpha}
\def \L {\Lambda}
\def \DD {{\cal D}}
\def \cN {{\cal N}}
\def \S {{\cal S}}

\def \W {{\cal W}}
\def \adss {$AdS_5 \times S^5$}
\def \ads {$AdS_5$}

\def \D {\Delta}
\def \Dd{{\bf D}}

\def \vt {\vartheta}
\def \lr { \lref}
\def \diag {{\rm diag}}
\def \z {\zeta}

\def \b {\beta}

\def \frac#1#2{{#1 \over #2}}
\def \R { R^{(2)}}

\def \l {\lambda}
\def \4p {{1\over 4 \pi }}
\def \8p {{1\over 8 \pi }}
\def \p {\phi}

\def \m {\mu }
\def \n {\nu}
\def \ep {\epsilon}
\def\g {\gamma}
\def \G {\Gamma}

\def \r {\rho}
\def \k {\kappa}
\def \d {\delta}

\def \s {\sigma}
\def \t {\theta}

\def \half {{1\over2}}

\def \fourth {{1\over 4}}

\def \const {{\rm const }}
\def\np { Nucl. Phys. }
\def \pl { Phys. Lett. }

\def \k {\kappa}

\def \g {\gamma}
\def \del {\partial}

\def \const {{\rm const}}

\def \b {\beta}
\def \r {\rho}
\def \s {\sigma}
\def \p {\phi}

\def \m {\mu}
\def \N {{\cal N}}
\def \n {\nu}
\def \vp {\varphi }

\def \t {\theta}

\def \G {{\Gamma}}
\def \td {\tilde }
\def \d {\delta}

\def \om {\omega}
\def \inv {^{-1}}
\def \ov {\over }
\def \four{{\textstyle{1\over 4}}}
\def \fourth{{{1\over 4}}}

\def \inf {{\infty}}
\def \zz {\td \zeta}


\lr\maldac{J.~Maldacena,
``The large-N limit of superconformal field theories and supergravity,''
Adv.\ Theor.\ Math.\ Phys.\ {\bf 2}, 231 (1998),
hep-th/9711200.
}

\lr\witt{E.~Witten,
``Anti-de Sitter space and holography,''
Adv.\ Theor.\ Math.\ Phys.\ {\bf 2}, 253 (1998),
hep-th/9802150.
}

\lr\gkp{S.S.~Gubser, I.R.~Klebanov and A.M.~Polyakov,
``Gauge theory correlators from non-critical string theory,''
Phys.\ Lett.\ {\bf B428}, 105 (1998),
hep-th/9802109.
}

\lr\gs {M.B. Green and J.H. Schwarz,
{Covariant description of superstrings},
 Phys. Lett. {\bf B136}, 367 (1984);
Nucl. Phys. {\bf B243}, 285 (1984). }

\lr\scht{A.S.~Schwarz and A.A.~Tseytlin,
``Dilaton shift under duality and torsion of elliptic complex,''
Nucl.\ Phys.\ {\bf B399}, 691 (1993),
hep-th/9210015.
}

\lr\alb {
A.S.~Schwarz,
``The Partition Function Of A Degenerate Functional,''
Commun.\ Math.\ Phys.\ {\bf 67}, 1 (1979).
}

\lr\mt {R.R.~Metsaev and A.A.~Tseytlin,
``Type IIB superstring action in $AdS_5 \times S^5$ background,''
Nucl.\ Phys.\ {\bf B533}, 109 (1998),
hep-th/9805028.
}

\lr\mye{E.~Myers,
``On The Interpretation Of The Energy Of The Vacuum As The Sum Over
Zero Point Energies,''
Phys.\ Rev.\ Lett.\ {\bf 59}, 165 (1987).
 }

\lr\kt {R.~Kallosh and A.A.~Tseytlin,
``Simplifying superstring action on $AdS_5 \times S^5$,''
JHEP {\bf 10}, 016 (1998),
hep-th/9808088. }

\lr\kr {
R.~Kallosh and J.~Rahmfeld,
``The GS string action on $AdS_5 \times S^5$,''
Phys.\ Lett.\ {\bf B443}, 143 (1998),
hep-th/9808038;
I.~Pesando,
``A kappa gauge fixed type IIB superstring action on
$AdS_5 \times S^5$,''
JHEP {\bf 9811}, 002 (1998),
hep-th/9808020.
}

\lr\mtt{ R.R. Metsaev and A.A. Tseytlin, unpublished;
 A.A. Tseytlin, talk at Strings' 98,
http://www.itp.ucsb.edu/online/strings98/tseytlin.
}

\lr\fgt{ S.~F\"orste, D.~Ghoshal and S.~Theisen,
``Stringy corrections to the Wilson loop in N = 4 super Yang-Mills theory,''
JHEP {\bf 08}, 013 (1999),
hep-th/9903042
}

\lr\ft {E.S. Fradkin and A.A. Tseytlin, {``Quantized string models,''}
Ann. of Phys. {\bf 143}, 413 (1982).
}

\lr\mal {J. Maldacena, {``Wilson loops in large N field theories,''}
Phys. Rev. Lett. {\bf 80}, 4859 (1998), {hep-th/9803002};
S.-J. Rey and J. Yee,
{``Macroscopic Strings as Heavy Quarks of Large N Gauge Theory and
Anti-de Sitter Supergravity,''}
{hep-th/9803001}.
}

\lr\alv{ D.~Friedan,
``Introduction To Polyakov's String Theory,''
{in Proc. of Summer School of Theoretical Physics: Recent
Advances in Field Theory and Statistical Mechanics, Les
Houches, France, Aug 2-Sep 10, 1982};
O.~Alvarez,
``Theory Of Strings With Boundaries: Fluctuations, Topology, And Quantum
Geometry,''
Nucl.\ Phys.\ {\bf B216}, 125 (1983);
H.~Luckock,
``Quantum Geometry Of Strings With Boundaries,''
Annals Phys.\  {\bf 194}, 113 (1989).

}

\lr\moor{
J.~Polchinski,
``Evaluation Of The One Loop String Path Integral,''
Commun.\ Math.\ Phys.\ {\bf 104}, 37 (1986);
G.~Moore and P.~Nelson,
``Absence Of Nonlocal Anomalies In The Polyakov String,''
Nucl.\ Phys.\ {\bf B266}, 58 (1986).
}

\lr\kalm{ R.~Kallosh and A.Y.~Morozov,
``Green-Schwarz Action And Loop Calculations For Superstring,''
Int.\ J.\ Mod.\ Phys.\ {\bf A3}, 1943 (1988).
}

\lr\mac {
P.B.~Gilkey,
``The Spectral Geometry Of A Riemannian Manifold,''
J.\ Diff.\ Geom.\ {\bf 10}, 601 (1975).
}

\lr\car{S.~Carlip,
``Heterotic String Path Integrals With The Green-Schwarz Covariant Action,''
Nucl.\ Phys.\ {\bf B284}, 365 (1987).
}

\lr\dgo {N.~Drukker, D.J.~Gross and H.~Ooguri,
``Wilson loops and minimal surfaces,''
Phys.\ Rev.\ {\bf D60}, 125006 (1999),
hep-th/9904191.
}

\lr\go{ J. Greensite and P. Olesen,
{``Remarks on the Heavy Quark Potential in the Supergravity Approach,''}
hep-th/9806235.
}

\lr\LU{L. Brink and H.B. Nielsen,
{A simple physical interpretation of the critical dimension of space time
in dual models},
 \pl {\bf B45}, 332 (1973);
M.~L\"uscher, K.~Symanzik and P.~Weisz,
``Anomalies Of The Free Loop Wave Equation In The WKB Approximation,''
Nucl.\ Phys.\ {\bf B173}, 365 (1980);
M. L\"uscher, {Symmetry breaking aspects of the roughening transition
in gauge theories}, \np {\bf B180}, 317 (1981);
O.~Alvarez,
``The Static Potential In String Models,''
Phys.\ Rev.\ {\bf D24}, 440 (1981).
}

\lr\dief{
K.~Dietz and T.~Filk,
``On The Renormalization Of String Functionals,''
Phys.\ Rev.\ {\bf D27}, 2944 (1983).
}


\lr\adtr{ I.~Pesando,
``The GS type IIB superstring action
on $AdS_3\times S^3\times T^4$,''
JHEP {\bf 02}, 007 (1999),
hep-th/9809145;
J.~Rahmfeld and A.~Rajaraman,
``The GS string action on $AdS_3\times S^3$
with Ramond-Ramond charge,''
Phys.\ Rev.\ {\bf D60}, 064014 (1999),
hep-th/9809164;
J.~Park and S.~Rey,
``Green-Schwarz superstring on $AdS_3 \times S^3$,''
JHEP {\bf 01}, 001 (1999),
hep-th/9812062.
}

\lr\wig{
P.B.~Wiegmann,
``Extrinsic Geometry Of Superstrings,''
Nucl.\ Phys.\ {\bf B323}, 330 (1989).
}

\lr\polya{
A.M. Polyakov, ``Two-dimensional Quantum gravity; Superconductivity
at high $T_c$,'' 1988 Les Houches School,
``Fields, Strings and Critical Phenomena,''
ed. by E. Br\' ezin and J. Zinn-Justin, North-Holland (1990),
p. 305.
}

\lr\lech{ K.~Lechner and M.~Tonin,
``The cancellation of worldsheet anomalies in the D=10 Green--Schwarz
heterotic string sigma--model,''
Nucl.\ Phys.\ {\bf B475}, 535 (1996),
hep-th/9603093.
}

\lr\oles{
P.~Olesen,
``Strings, Tachyons And Deconfinement,''
Phys.\ Lett.\ {\bf 160B}, 408 (1985).
}

\lr\pora{
M.~Porrati and P.~van Nieuwenhuizen,
``Absence of world sheet and space-time anomalies in the semicovariantly
quantized heterotic string,''
Phys.\ Lett.\ {\bf B273}, 47 (1991).
}

\lr\naiv{U.~Kraemmer and A.~Rebhan,
``Anomalous Anomalies In The Carlip-Kallosh Quantization Of The
Green-Schwarz Superstring,''
Phys.\ Lett.\ {\bf B236}, 255 (1990).
F.~Bastianelli, P.~van Nieuwenhuizen and A.~Van Proeyen,
``Superstring anomalies in the semilight cone gauge,''
Phys.\ Lett.\ {\bf B253}, 67 (1991).
}

\lr\kav{
A.R.~Kavalov and A.G.~Sedrakian,
``Quantum Geometry Of Covariant Superstring With N=1 Global Supersymmetry,''
Phys.\ Lett.\ {\bf 182B}, 33 (1986).
}

\lr\sprad{
J.~Michelson and M.~Spradlin,
``Supergravity spectrum on $AdS_2 \times S^2$,''
JHEP {\bf 09}, 029 (1999),
hep-th/9906056.
}

\lr\ivan{
E.A.~Ivanov and A.S.~Sorin,
``Wess-Zumino Model As Linear Sigma Model Of Spontaneously Broken
Conformal And Osp(1,4) Supersymmetries,''
Sov.\ J.\ Nucl.\ Phys.\ {\bf 30}, 440 (1979);
``Superfield Formulation Of Osp(1,4) Supersymmetry,''
J.\ Phys.\ {\bf A13} (1980) 1159.
}

\lr\cahi{
R.~Camporesi and A.~Higuchi,
``Stress energy tensors in anti-de Sitter space-time,''
Phys.\ Rev.\ {\bf D45}, 3591 (1992).
}

\lr\sedr{
A.R.~Kavalov, I.K.~Kostov and A.G.~Sedrakian,
``Dirac And Weyl Fermion Dynamics On Two-Dimensional Surface,''
Phys.\ Lett.\ {\bf B175}, 331 (1986);
A.G.~Sedrakian and R.~Stora,
``Dirac And Weyl Fermions Coupled To Two-Dimensional Surfaces: Determinants,''
Phys.\ Lett.\ {\bf 188B}, 442 (1987);
D.R.~Karakhanian and A.G.~Sedrakian,
``Heterotic string bosonization in a space of
 arbitrary dimension",
preprint YERPHI-1127(4)-89.
}

\lr\kara{
D.R.~Karakhanian,
``Induced Dirac Operator And Smooth Manifold Geometry,''
preprint YERPHI-1246-32-90 (1990).
}

\lr\leu{ F.~Langouche and H.~Leutwyler,
``Two-Dimensional Fermion Determinants As Wess-Zumino Actions,''
Phys.\ Lett.\ {\bf 195B}, 56 (1987);
``Weyl Fermions On Strings Embedded In Three-Dimensions,''
Z.\ Phys.\ {\bf C36}, 473 (1987);
``Anomalies Generated By Extrinsic Curvature,''
Z.\ Phys.\ {\bf C36}, 479 (1987).
}

\lr\pow{
A.M.~Polyakov and P.B.~Wiegmann,
``Theory of nonabelian Goldstone bosons in two dimensions,''
Phys.\ Lett.\ {\bf 131B}, 121 (1983);
``Goldstone Fields In Two-Dimensions With Multivalued Actions,''
Phys.\ Lett.\ {\bf 141B}, 223 (1984).
}

\lr\polu{A.M. Polyakov, unpublished (1986).}

\lr \FT{
E.S.~Fradkin and A.A.~Tseytlin,
``Quantum String Theory Effective Action,''
Nucl.\ Phys.\ {\bf B261}, 1 (1985).
}

\lr\campo{
R.~Camporesi,
``Zeta function regularization of one loop effective potentials in anti-de
Sitter space-time,''
Phys.\ Rev.\ {\bf D43}, 3958 (1991);
R.~Camporesi and A.~Higuchi
``Arbitrary-spin effective potentials in anti-de Sitter spacetime,''
Phys.\ Rev.\ {\bf D47}, 3339 (1993).
}

\lr\camph{
R.~Camporesi and A.~Higuchi,
``Spectral functions and zeta functions in hyperbolic spaces,''
J.\ Math.\ Phys.\ {\bf 35}, 4217 (1994).
}

\lr\dcfm{
D.~Berenstein, R.~Corrado, W.~Fischler and J.~Maldacena,
``The operator product expansion for Wilson loops and surfaces in the large
N limit,''
Phys.\ Rev.\ {\bf D59}, 105023 (1999),
hep-th/9809188.
}

\lr\burg{
C.P.~Burgess,
``Supersymmetry Breaking And Vacuum Energy On Anti-De Sitter Space,''
Nucl.\ Phys.\ {\bf B259}, 473 (1985).
}

\lr\barf{
W.A.~Bardeen and D.Z.~Freedman,
``On The Energy Crisis In Anti-De Sitter Supersymmetry,''
Nucl.\ Phys.\ {\bf B253}, 635 (1985).
}

\lr\iopot{
T.~Inami and H.~Ooguri,
``Dynamical Breakdown Of Supersymmetry In Two-Dimensional Anti-De Sitter
Space,''
Nucl.\ Phys.\ {\bf B273}, 487 (1986).
}

\lr\ald{
B.~Allen and S.~Davis,
``Vacuum Energy In Gauged Extended Supergravity,''
Phys.\ Lett.\ {\bf 124B}, 353 (1983).
}

\lr \poll{A.M. Polyakov, unpublished.}

\lr\ina{
T.~Inami and H.~Ooguri,
``One Loop Effective Potential In Anti-De Sitter Space,''
Prog.\ Theor.\ Phys.\ {\bf 73}, 1051 (1985).
}

\lr\sakpot{
N.~Sakai and Y.~Tanii,
``Effective Potential In Two-Dimensional Anti-De Sitter Space,''
Nucl.\ Phys.\ {\bf B255}, 401 (1985).
}

\lr\breit{
P.~Breitenlohner and D.Z.~Freedman,
``Stability In Gauged Extended Supergravity,''
Ann.\ Phys.\ {\bf 144}, 249 (1982).
}

\lr\avi{
S.J.~Avis, C.J.~Isham and D.~Storey,
``Quantum Field Theory In Anti-De Sitter Space-Time,''
Phys.\ Rev.\ {\bf D18}, 3565 (1978).
}

\lr\gibbo{
C.J.~Burges, D.Z.~Freedman, S.~Davis and G.W.~Gibbons,
``Supersymmetry In Anti-De Sitter Space,''
Ann.\ Phys.\ {\bf 167}, 285 (1986).
}

\lr\bul{
C.P.~Burgess and C.A.~Lutken,
``Propagators And Effective Potentials In Anti-De Sitter Space,''
Phys.\ Lett.\ {\bf 153B}, 137 (1985).
}

\lr\sakone{
N.~Sakai and Y.~Tanii,
``Supersymmetry And Vacuum Energy In Anti-De Sitter Space,''
Phys.\ Lett.\ {\bf 146B}, 38 (1984).
}

\lr\stpo{
J.~Polchinski and A.~Strominger,
``Effective string theory,''
Phys.\ Rev.\ Lett.\ {\bf 67}, 1681 (1991).
}

\lr \saktan{
N.~Sakai and Y.~Tanii,
``Supersymmetry In Two-Dimensional Anti-De Sitter Space,''
Nucl.\ Phys.\ {\bf B258}, 661 (1985).
}

\lr\kinar{Y.~Kinar, E.~Schreiber, J.~Sonnenschein and N.~Weiss,
``Quantum fluctuations of Wilson loops from string models,''
hep-th/9911123.
}

\lr\naik{S.~Naik,
``Improved heavy quark potential at finite temperature from anti-de
Sitter supergravity,''
Phys.\ Lett.\ {\bf B464}, 73 (1999),
hep-th/9904147.
}

\lr\polyakov{A.M.~Polyakov,
``Quantum geometry of bosonic strings,''
Phys.\ Lett.\ {\bf B103}, 207 (1981);
``Quantum geometry of fermionic strings,''
Phys.\ Lett.\ {\bf B103}, 211 (1981).
}

\lr \fini{
S.M.~Christensen, M.J.~Duff, G.W.~Gibbons and M.~Rocek,
``Vanishing One Loop Beta Function In Gauged $N > 4$ Supergravity,''
Phys.\ Rev.\ Lett.\ {\bf 45}, 161 (1980).
}

\Title{\vbox
{\baselineskip 12pt{\hbox{NSF-ITP-99-146 }}
{\hbox{OHSTPY-HEP-T-99-029}}
{\hbox{hep-th/0001204}}
\vskip-.3in
}}
{\vbox{\centerline{ Green-Schwarz String in
 $AdS_5 \times S^5$:}
\vskip2pt\centerline{ Semiclassical Partition Function}}}
\centerline {Nadav Drukker,$^{1,2}$
David J. Gross,$^1$
and Arkady A. Tseytlin$^3$
\footnote{$^\star$} { Also at Lebedev
Physics Institute, Moscow and Imperial College, London. }
}
\baselineskip=12 pt
\bigskip
\centerline {$^1$ Institute for Theoretical Physics}
\centerline {University of California }
\centerline {Santa Barbara CA, 93106-4030}
\medskip
\centerline {$^2$ Department of Physics}
\centerline {Princeton University}
\centerline {Princeton NJ, 08544}
\medskip
\centerline {$^3$ Department of Physics}
\centerline {The Ohio State University}
\centerline {Columbus OH, 43210-1106}
\medskip
\centerline {\tt email: drukker,
gross@itp.ucsb.edu,\
tseytlin@mps.ohio-state.edu}

\baselineskip=16pt plus 2pt minus 2pt
\vskip .3in
\centerline{\bf Abstract}
\smallskip
\centerline{\vbox{\hsize=4.5in
A systematic approach to the study of semiclassical fluctuations
of strings in \adss\ based on the Green-Schwarz formalism
is developed. We show that the string partition
function is well defined and finite. Issues related to different
gauge choices are clarified.
We consider explicitly several cases of
classical string solutions with the world surface ending
on a line, on a circle or on two lines on the boundary of $AdS$.
The first example is a BPS object and the partition function is
one. In the third example
the determinants we derive should give the first corrections to the
Wilson loop expectation value in the strong coupling expansion of
the $\cN=4$ SYM theory at large $N$.
}}

\Date {January 2000}

\newsec{Introduction}

The duality between \adss\ and $\cN=4$ super Yang-Mills theory in
four dimensions is the best studied example of the $AdS$/CFT
correspondence \refs{\maldac,\gkp,\witt}. This duality allows the
calculation of gauge theory observables at large $N$ and large
't Hooft coupling from perturbative supergravity or string theory.
In particular, Wilson loops are described by classical strings
that end at the boundary of $AdS$ \mal.

To extend this duality beyond the supergravity limit it is necessary
to learn how to handle strings on this space. Because the background
includes a Ramond-Ramond 5-form flux, it is difficult to use the
RNS formalism to quantize strings in this geometry. Therefore, one
is led to use the Green-Schwarz (GS) action. The first step in this
direction was the construction of the classical GS action for strings
on this background \mt.

Even in flat space the GS action is hard to quantize, except in
the light cone gauge. However, this action is perfectly applicable
to the perturbative analysis of quantum corrections around a
non-trivial ``long string" classical solution (assuming the
classical bosonic background makes the fermionic kinetic term
well-defined). This strategy can be applied in either flat or
curved space, and, in particular, is well suited for strings
in \adss\ where there is a natural static solution \refs{\mal}
to expand about.

The string 2-d loop expansion in \adss\ is an expansion
in powers of $\a'/R^2 = \l^{-1/2}$, where $R$ is the radius
parameter of \adss\ and $\lambda$ is the 't Hooft coupling:
the leading term coming from the classical action is
proportional to $\sqrt \l$, the 1-loop correction is just a
number, the 2-loop correction will be multiplied by
$\a'/R^2 = \l^{-1/2}$, etc.

The main goal of this paper is to develop technical tools
necessary to do calculations of quantum string corrections
in \adss, at least in the one-loop approximation. This is
an important step in the extension of the AdS/CFT
correspondence beyond the classical level. Our main motivation is
to find the quantum string correction to the Wilson loop
expectation value, in particular, the first sub-leading
(i.e. $\l$-independent) correction to the quark anti-quark
potential.

This problem was first addressed in \kt, where the relevant
fermionic operator coming from GS action was derived. An
important next step was made in \fgt, where the partition
function was expressed in terms of operators defined with
respect to the induced 2-d geometry. Refs.
\refs{\go,\naik,\kinar} also discussed corrections to the
quark anti-quark potential in \adss\ and in other related
geometries. However, all these previous attempts were
incomplete as they encountered problems with divergences,
gauge fixing, and other subtleties.

Our aim is to clarify some of these issues and to set up
a consistent framework for performing the semiclassical
calculations for the GS string in a curved target space.
In particular, we shall explain how the divergence
proportional to the world sheet curvature $\R$ found in
\fgt\ is canceled (the cancellation of this divergence in the
one-loop approximation in curved target space is
essentially the same as in flat space). We will also
explain the close relation between the fermionic operators
in \kt\ and in \fgt\ (they correspond to two choices of
$\k$-symmetry gauge).

The paper is organized as follows.

We start with some general comments about the Green-Schwarz
action in flat space, and, in particular, how to use it
to calculate quantum corrections to a classical solution.
This involves gauge fixing, and determining the measure
in the path integral. We will find it most reliable to use
conformal gauge, where the path integral measure is best
understood \refs{\polyakov,\alv}. Since the theory is critical,
the conformal anomalies and, therefore, the 2-d divergences,
cancel out. The same mechanism is responsible for the
cancellation of leading-order (one-loop) divergences
(that are proportional to $\R$) in curved space as well.

In Section 3 we turn to strings on \adss. We review the
 corresponding Brink-DiVecchia-Howe-Polyakov type
GS action and explain how to evaluate the quadratic
fluctuations around a classical solution. We also comment
on the approach based on the Nambu-Goto type
action in the static gauge. With careful account of ghosts
(and path integral measures) the two approaches should give
the same results.

We show that as in \refs{\sedr,\leu,\wig,\kara,\lech} a local
Lorentz rotation of GS spinors allows one to systematically
 transform the quadratic fermionic term in the GS action into
the action for a set of 2-d fermions. The problem of
computing the partition function is then reduced to the
evaluation of determinants of some bosonic and fermionic
operators on a 2-d world sheet with an induced metric that is
asymptotic to $AdS_2$.

We study three special examples.
In Section 4 we consider a string world surface that ends on
a single straight line at the boundary of $AdS_5$. The induced
metric on the world surface is that of $AdS_2$ and the quantum
fluctuation fields fit nicely into supersymmetry multiplets
on that space. We compute the corresponding vacuum energy and
show that it vanishes using a $\zeta$-function regularization.
The vacuum energy is related to the partition function by a
conformal anomaly. Using that we show that the partition
function is equal to one.

Another case which leads again to $AdS_2$ for the induced 2-d
geometry is a circular Wilson loop, which we study in Section 5.
We comment on the difference between the circular and the straight
line cases.

In Section 6 we turn to the case of most interest, the
surface corresponding to the quark -- anti-quark system.
Here the induced geometry is more complicated
 but is still asymptotically $AdS_2$.
We derive the general expression for the partition function
(demonstrating in the process the equivalence of the two
$\k$-symmetry gauges $\t^1=\t^2$ and $\t^1=i\G_4 \t^2$)
and discuss evaluation of the numerical coefficient
in the corresponding one-loop correction to the $1/L$ potential
using a crude approximation to the geometry.

We summarize our results in Section 7.

Some general remarks and explicit calculations are given in
Appendices. In Appendix A we review how a determinant of a
Laplace operator changes under rescaling of the measure of
the fields. The resulting general relations are useful in
computing various contributions to the partition function.

In Appendix B we present two different calculations
of the partition function in the case of $AdS_2$ as
the induced geometry.

Some comments about the fermionic 2-d determinants
related to GS action are given in Appendix C.

In Appendix {D} we point out that the expression for the
superstring partition function in $AdS_3\times S^3$ with RR
2-form background is very similar to the one in the
$AdS_5 \times S^5$ case.

Below we shall use the following notation:
$i,j, ...=0,1$ and $\a,\b,...=0,1$ will denote 2-d
world and tangent space indices;
$a,b,...=0,...,4$ and $p,q,...=1,...,5$
will be the tangent space indices of $AdS_5$ and $S^5$;
$\hat a =0,1,...,9$ will be the tangent space indices
of the 10-d space-time.

\newsec{Green-Schwarz action in flat space}

Before plunging into discussion
of strings in curved target space,
it is useful to clarify
several general points about the GS action.
The flat space GS action of type IIB theory is \gs\
\eqn\gre{\eqalign{
S_{\rm flat}=
 {1\over 2\pi \alpha'} \int d^2 \sigma \bigg[
&\,- \ha \sqrt{g} g^{ij} \eta_{\hat a \hat b} \left(\del_i x^{\hat a} -
i \bar \t^I \G^{\hat a} \del_i \t^I\right)
\left(\del_j x^{\hat b} -
i \bar \t^J \G^{\hat b} \del_j \t^J\right)
\cr&
 - i \ep^{ij} s^{IJ} \bar \t^I \G_{\hat a} \del_j \t^J
\left( \del_i x^{\hat a} - \ha i
\bar \t^K \G^{\hat a} \del_i \t^K\right)
 \bigg],
}}
where $\hat a = 0,1,...,9$, \ $s^{IJ}$ is defined by
$s^{11}=-s^{22}=1$, $s^{12}=s^{21}=0$,
$g_{ij}$ ($i,j=0,1$) is a world-sheet metric with signature
$(-+)$, $g=-\det g_{ij}$, and $\t^I$ are two left 10-d Majorana Weyl
spinors.

This action can be considered in either the Polyakov
form, with independent 2-d metric (which can be quantized in
the conformal gauge) or in the Nambu form, with the induced metric
(which can be quantized in the static gauge). When doing semiclassical
expansion near a ``long string" configuration it may seem natural
to use the Nambu formulation, choosing a static gauge.
However, the meaning of conformal invariance conditions
and the definition of the path integral measure are
clear only in the Polyakov formulation. In that case,
the 2-d metric, which at the classical level is proportional
to the induced metric, should be treated as independent
of the coordinates in checking the conformal invariance
constraints on the background target space fields
(in particular, in proving that conformal anomalies cancel
in flat $D=10$ space). In the leading 1-loop approximation
the Polyakov and Nambu formulations are expected
to produce equivalent expressions for the partition
function. However, the precise way the divergences cancel
may become rather obscure once one sets the metric to be
equal to the induced metric, since $\int \R$ and
$\int \del x \del x$ divergences may get mixed up. In particular,
the $\int \R$ divergences become equivalent to total derivative
contributions to $x$-dependent divergences and reduce to boundary
terms (which may eventually cancel against boundary counterterms).

Another important point concerns the distinction between the
fermionic kinetic term for GS fermions and for standard 2-d
Dirac fermions. As was observed in \refs{\wig,\lech} (see also
\refs{\polu,\sedr,\kara,\leu,\polya}) in the case of a flat
target space, one may perform a local target space rotation
that transforms the quadratic GS fermion term into the 2-d
fermion kinetic term. The resulting Jacobian (see, in
particular, \leu) depends on the 2-d metric and its contribution
explains why the conformal anomaly of a GS fermion is 4 times
bigger than that of a 2-d fermion \kalm\ (which is crucial
for understanding how conformal anomalies cancel in $D=10$
GS string). Similar remarks apply in the case of curved
target spaces. As we shall explicitly discuss below, in
some simple cases (like the straight string in \adss) the
quadratic part of the GS action has already the 2-d fermion
form with respect to the curved geometry of induced metric.
In other cases one must perform a rotation to express the
action in the 2-d fermion form. In the Polyakov formulation
with independent 2-d metric, the Jacobian of this must be
taken into account for consistent cancellation of conformal
anomalies. The contribution of this Jacobian may be non-trivial
also in the Nambu formulation where it may depend on the
$x$-background.

\subsec{\bf Quadratic fluctuations near a classical solution}

The Green-Schwarz action \gre\ is not quadratic
in fermions, and is difficult to quantize.
One standard way to proceed is to choose a light cone gauge.
Alternatively, one may resort to a perturbative expansion
in powers of $\a'$ near
a particular classical solution.
Since the latter strategy is the only one available
in the curved $AdS_5 \times S^5$ case, we shall employ it below.
We concentrate on the one-loop approximation,
i.e. on the leading quantum correction to the partition
function of the GS string action expanded
near a classical solution.

With a suitable choice of coordinates, we can
write the ``long string" classical solution as
\eqn\flalst{
x^0 = \sigma^0\ ,
\qquad
x^1 = \sigma^1\ .
}
The bosonic part of the action \gre\ is simply the Polyakov
action, and it can be quantized in the conformal gauge
$\sqrt g g^{ij} =\d^{ij}$. This results in 10 massless
world-sheet scalars and two ghosts.

Alternatively, one could start with the Nambu form of the
action (i.e. first solve for $g_{ij}$ and then quantize
the theory). In that case we can again expand near \flalst\
and choose the static gauge, i.e. eliminate the fluctuations
in $(0,1)$ directions. Then we are left with just eight
transverse scalars.

The quadratic term in the fermionic part of the GS action is
\eqn\uio{ S_{\rm 2F}={1\over 2\pi \alpha'}\int d^2 \sigma\
L_{\rm 2 F} =
{i\over 2\pi \alpha'} \int d^2 \sigma
\left( \sqrt{g} g^{ij}\d^{IJ} - \ep^{ij} s^{IJ}\right)
\bar \t^I \r_i \del_j \t^J,
}
where $g_{ij}$ can be set equal to $\eta_{ij}$ in the conformal
gauge. $\r_i$ is the projection of the 10-d Dirac matrices on
the world sheet
\eqn\utr{
\r_i \equiv \G_{\hat a} \del_i x^{\hat a} = \G_i\,,
}
where the last equality is true for the classical solution
\flalst. Then
\eqn\ferm{
L_{\rm 2 F}
= i \bar\t^1 \G^+ \del_+ \t^1
+ i\bar\t^2 \G^- \del_- \t^2 \,.
}
This fermionic action is obviously invariant under
$\delta \t^1 = \G^+ \k^1$, $\delta \t^2 = \G^- \k^2$
which is just the leading-order term in the $\k$-symmetry
transformation rules
\eqn\kap{ \delta_\k \t^I = \r_i \k^{i I} + ...\ ,
\qquad
\frac{1}{\sqrt{g}}\epsilon^{ij}\kappa_j^1=-\kappa^{i1},
\qquad
\frac{1}{\sqrt{g}}\epsilon^{ij}\kappa_j^2=\kappa^{i2}.
}
Since we can represent the (left subspace, 16-component)
10-d Dirac matrices in the form
$\G_{i} = \tau_i \times I_8 $, where
$\tau_i$ are $ 2\times 2$ Dirac matrices and $I_8$ is the
$8\times 8$ unit matrix, the
meaning of the $\k$-symmetry transformations in the present case
is simply that $\t^1$ corresponds to eight left 2-d spinors
and $\t^2$ to eight right 2-d spinors.

A natural way to fix $\k$-symmetry is to set\foot{This gauge
(considered also in \kav) is possible only in type IIB theory
where the two spinors have the same chirality. This gauge is also
natural in connection with the open string theory---in type I
theory $\t^1=\t^2$ at the boundary of world sheet.
}
\eqn\asq{ \t^1 =\t^2 \equiv \t. }
The remaining degrees of freedom are then 8 real 2-d spinors
represented by 16-component left 10-d MW spinor $\t$.

The global part of the local $\k$-symmetry transformation of $x^\m$,
that is preserved by the above gauge choice, may then be interpreted
as the effective 2-d supersymmetry of the resulting quadratic action,
$\delta x^k = \bar \t^I \G^k \delta_\k \t^I$.
This is similar to what happens in the light cone gauge. As we shall
see, this simple picture has a direct counterpart in curved case.

\subsec{\bf Conformal invariance of GS string in flat space}

The proof that the fermionic RNS string is conformally invariant
at the quantum level \polyakov\ is based on adding together the
central charges of all the fields: 1 for each scalar boson,
$1/2$ for each Majorana 2-d fermion, $-26$ for the
conformal ghosts and $11$ for the superconformal ghosts.

Since conformal anomalies are associated with UV divergences,
it is not surprising that the same counting is responsible
for the cancellation of logarithmic divergences in the
properly defined string partition function on a 2-d surface
with any number of holes and handles \alv. This is obvious
for the scalar and fermion determinants. For the ghosts,
the essential extra ingredient is the need to take into account
some global factors in the gauge group measure associated with
conformal Killing vectors and/or Teichm\"uller moduli.
Then the logarithmic divergences are again
proportional to the total central charge times the Euler number of the
Riemann surface and cancel out if $D=10$ (or $D=26$ in the
bosonic string case).

The counting for the GS string is different.
We describe here only the one-loop approximation.
To discuss the cancellation of conformal anomaly
in GS string we need to keep
the dependence on a generic fiducial metric $g_{ij}$
in the action \uio\ and in the norms of the fields
($\int d^2 \s\sqrt g \bar \t \t$, etc.).
Naively, after gauge fixing one gets 10 scalars,
8 Majorana 2-d fermions and the bosonic conformal ghosts.
The naive counting would give $10\times1+8\times\ha-26=-12$.
But, in fact, the GS fermionic action depends on the 2-d metric,
not as in the case of the standard action for a 2-d spinor
(i.e. not through $\sqrt g\ e^i_\a$ where $e^\a_i$ is a
zweibein), but rather as a 2-d scalar action
(i.e. through $\sqrt g\ g^{ij}$).
In the conformal gauge $g_{ij} =e^{2\rho} \eta_{ij},
\ e^\a_i = e^{\rho }\d^\a_i $ that effectively results in
the replacement of $\rho$ by $2\rho$ in the conformal anomaly
term $(\int \del \rho \bar \del \rho)$ for a 2-d
spinor, giving {\it four} times bigger a result \kalm.

Hence the contribution of each of 8 species of GS fermions to
the divergence is effectively as of 4 \ 2-d spinors.\foot{In
more detail, the action for a 2-d spinor is
$\int d^2\s \sqrt g g^{ij} \bar \psi e^\a_i \tau_\a \del_j \psi$,
while for the fermions in the GS action \uio\
the world sheet combination $\tau_i=e_i^\a\tau_\a$ is
replaced by the target space one, $\r_i=\del_i x^{\hat a}
\Gamma_{\hat a}$. In certain cases the two might be equal,
but they do not behave the same way under the conformal
transformations of the world-sheet metric. The GS fermions
$\theta$ are world sheet scalars, so their natural measure is
$\|\t\|^2 = \int d^2 \s \sqrt g \bar \t\t$. In the conformal
gauge ($g_{ij} = \sqrt g \d_{ij}$) the GS fermionic action is
$\int d^2 \s \bar \t \rho_\a \del_\a \t$. Because of the
normalization of the $\theta$'s, after squaring the fermionic
operator we get
${ 1 \ov \sqrt g} \del { 1 \ov \sqrt g} \bar \del$.
In the case of the 2-d spinors in the conformal gauge the zweibein
contributes to the scaling of the action
$\int d^2\s (\sqrt g )^{1/2} \bar
\psi \tau_\a \del_\a \psi$,
\ $\|\psi\|^2 =\int d^2 \s \sqrt g \bar \psi \psi$.
If we rescale $\psi$ to make the action $g$-independent as in $\t$
case we get
$\int d^2\s \bar \psi \tau_\a \del_\a \psi$,
$\|\psi\|^2 =\int d^2 \s (\sqrt g)^{1/2} \bar \psi \psi$.
The difference compared to $\t$ is now only in the norm.
The corresponding 2-nd order operator is
${ 1 \ov (\sqrt g)^{1/2}} \del { 1 \ov (\sqrt g)^{1/2}} \bar \del$.
It is the difference in the measure factor that now leads to
different anomalies: for the operator $f \del f \bar \del$
the conformal anomaly in the partition function is
$\exp ( - { 1 \ov 48 \pi} \int d^2\s | \del \ln f|^2 )$,
so that the difference between the ``scalar'' and
``2-d spinor'' descriptions produces indeed the factor of 4 in
the conformal anomaly.
}
Then the count of anomalies in the GS string goes as follows
\eqn\wrong{
10-26 +8\times4\times \ha = 0
\,.}
Essentially equivalent arguments (based on separating the metric
dependence in a WZ type Jacobian contribution due to a rotation
of spinors) which explains why conformal anomaly cancels in
$D=10$ in GS string were given in \refs{\wig,\lech,\leu}
(see also Appendix C).

Since in a covariant regularization the cutoff is coupled
to the conformal factor, the cancellation of conformal anomalies
should imply also the cancellation of the UV divergences.

\subsec{\bf Cancellation of quantum correction to
straight string configuration}

A natural classical ``long string'' solution in flat space
is \flalst, or restoring the dimensional parameters,
\eqn\fllst{
x^0 = T \tau,\ \ \ \ \
\qquad
x^1 = L \s,\ \ \ \ \ \
\qquad
\s\in (-\ha, \ha).
}
The fluctuations of the $d-2=8$ transverse bosonic coordinates
(which are periodic in $\s^0$ and Dirichlet in $\s^1$ directions)
give for $T\to \infty$ \refs{\LU}
\eqn\aca{
W = - \ln Z = (d-2) W_0 \,,
\qquad
W_0 = \ha [ \log \det (- \partial^2)]_{T\to \infty}
= - {\pi \ov 24 }\ {T \ov L}\,
.}
In the flat-space superstring case this contribution is canceled
by the contribution of the fermionic determinant: the total
effective number of transverse world-sheet degrees of freedom is
equal to zero (as in any flat-space string theory without tachyons
\oles) because of the effective 2-d supersymmetry present after
choosing $\t^1=\t^2$ and expanding to quadratic order near
\fllst. Indeed, the induced metric and zweibein are flat, and
thus, apart from the subtlety with cancellation of conformal
anomalies discussed above, the GS fermionic determinant is the
same as for eight 2-d spinors.

\newsec{Quadratic fluctuations of superstring in \adss}

We now turn to the discussion of the one-loop approximation
to the partition function of GS superstring in \adss. We
start with the Polyakov form of GS action in conformal gauge,
expand near a general classical solution,
and explicitly check conformal invariance to 1-loop order.
We shall also comment on the result obtained
by starting with the Nambu-type action and using static gauge.
In the following sections we give examples of
particular symmetric solutions.

In the context of the $AdS$/CFT duality, expansion about
classical solutions of the string action, namely minimal
surfaces, corresponds to computing expectation values
of Wilson loop operators in the dual gauge theory \mal.
The expectation value of the Wilson loop is given by
\eqn\wilvev{
\vev{W} = \int [dx][d\theta][d g]\ e^{-S}
, }
where $S$ is the string action in \adss\ and
the path integral is over all embeddings of the string into \adss\
with proper boundary conditions (the string world
surface should end along the loop at the boundary
of $AdS_5$ \refs{\mal,\dgo}). Here we assumed the Polyakov form,
where, in general, one is to integrate over the moduli of
2-d metrics.

\subsec{\bf The action}

The bosonic part of the action for a string in \adss\ is\foot{For a
string representing a Wilson loop of SYM theory
and ending at the boundary the classical
action is actually a particular Legendre transform of the area \dgo,
but that does not affect the discussion of quantum fluctuations.
}
\eqn\acto{
S_{\rm B}=
 {R^2\over 4\pi \alpha'} \int d^2 \sigma
\sqrt{g} g^{ij} G_{\mu\nu}(x)\del_i x^\mu \del_j x^\nu\
.}
We have removed the dependence on the $AdS$ scale
$R$ \ ($R^4 = 4 \pi \a'^2 g_s N$) from the space-time metric
$G_{\mu\nu}$ \
($m=1,\dots,4$)
\eqn\adsmet{
ds^2={1\over w^2}\left(dw^2+dx^mdx^m\right)+d\Omega_5^2
\,.}
The leading behavior at large 't Hooft coupling $\l$ is the exponent
of the classical action, which is proportional to ${R^2\ov \alpha'}
= \sqrt\lambda$. The string expansion is in inverse powers of
$\sqrt\lambda$. In most of the paper we set $R=1$, but it is
easy to restore the dependence on $R$ when necessary.

The structure of the full covariant GS string action in \adss\
is rather complicated \mt, but the part quadratic in $\t^I$ is
simple and is a direct generalization of the quadratic term in
the flat-space GS action \ferm\
\eqn\frmo{
S_{\rm 2F} = {i \ov 2 \pi \a'} \int d^2 \s
( \sqrt{g} g^{ij}\d^{IJ} - \ep^{ij} s^{IJ} )
\bar \t^I \r_i D_j \t^J\ .
}
Here $\r_i$ are again projections of the 10-d Dirac matrices,
\eqn\trw{
\r_i \equiv \G_{\hat a} E^{\hat a}_\m \del_i x^\m =
( \G_a E^a_\m +\G_p E^p_\m )
 \del_i x^\m \ ,
}
and $E^{\hat a}_\m$ is the vielbein of the 10-d target space metric,
$G_{\m\n} = E^{\hat a }_\m E^{\hat b}_\n \eta _{\hat a \hat b}$.
The covariant derivative $D_i$ is the projection of the
10-d derivative
$D_{ \mu}=\del_{ \mu}
+ \fourth\Omega^{\hat a\hat b}_{ \m} \Gamma_{\hat a\hat b}
 - { 1 \ov 8 \cdot 5!}
 \G^{ \m_1...\m_5} \G_{ \m}\ e^\phi F_{ \m_1... \m_5}$
($\Omega^{\hat a\hat b}_\m$ is the spin connection and
$F_{ \m_1... \m_5}$ the RR 5-form potential)
which appears, e.g., in the Killing spinor equation of type
IIB supergravity. It has the following explicit form
\mt\foot{The 10-d Dirac matrices are split in the `5+5' way,
$\Gamma^a= \gamma^a \times I_4 \times \s_1$,
$\Gamma^{p}= I_4 \times \gamma^{p} \times \s_2,$
where $\s_k$ are Pauli matrices and
$\g^a$, $\g^{p}$ are $4\times 4$ matrices
(corresponding to tangent spaces of $AdS_5$ and $S^5$ factors)
and $I_4$ is $4\times 4$
unit matrix (see \mt\ for details on notation;
we use index $p=1,...,5$ instead of $a'$ in \mt).
A $D=10$ positive chirality 32-component spinor $ \Psi$ is decomposed
 as follows:
$\Psi =\psi\times \psi^\prime \times \left({1\atop 0}\right)$.
In equations written in the 32-component spinor form,
$\theta^I $ stands for two positive chirality spinors
$\left({\theta^I\atop 0}\right) $,
where $\theta^I$ are 16-component spinors used below.
The Majorana condition $\bar \Psi \equiv \Psi^\dagger \G^0=
\Psi^T \CC$, \ $\CC\equiv C \times C' \times i \s_2$,
then takes the form
$\bar \theta_{\a\a'I} \equiv (\theta^{I\b\b'} )^\dagger (\g^0)^\b_\a
\delta^{\b'}_{\a'} = \theta^{I\b\b'} C_{\b\a} C'_{\b'\a'}$.
\ Here $C$ and $C'$ are the charge conjugation matrices of $so(4,1)$
and $so(5)$ used to raise and lower spinor indices.
Note that $C\times C'$ is symmetric, i.e. $\bar \t \t=0$.
In the expressions below $\t^I$ may be thought of as 5-d spinors
with an extra `spectator' 5-d spinor index, and we shall assume that
$\gamma^a$ and $\gamma^p$ stand for $\gamma^a \times I_4$ and
$I_4\times \gamma^p$.
$C\gamma^{a_1\ldots a_n}$ are symmetric (antisymmetric)
for $n=2,3$ mod 4 ($n=0,1$ mod 4). The same properties are valid
for $C^\prime\gamma^{p_1\ldots p_n}$.
We also assume that $\g_0^\dagger = -\g_0$, $\g_m^\dagger =\g_m$\
$(m=1,2,3,4)$.
}
\eqn\cov{
D_i\t^I \equiv \left(\delta^{IJ} {\DD_i}
- \half {i} \epsilon^{IJ} \tilde\r_i\right) \t^J,
\qquad
{\DD_i}= \del_i
+\fourth \del_i x^\m \Omega^{\hat a\hat b}_\m\Gamma_{\hat a\hat b}\,
, }
where the $\tilde\r_i$ term originates from the coupling
 to the RR field strength,
\eqn\nott{
\tilde\r_i \equiv \left(\G_a E^a_\m +i\G_p E^p_\m\right)
 \del_i x^\m\ . }
Note that $\tilde\r_i$ is not identical to $\rho_i$, unless
one is expanding near a classical solution that is constant
on $S^5$.

In general, there is a factor of $R\inv$ in front of the
`mass term', so that this term disappears in the flat-space
limit.

\subsec{\bf Expanding about a classical solution}

We first consider the bosonic sector.
Expand the Polyakov action \acto\ about a classical solution
\eqn\smlfl{
x^\m\to \bar x^\m + \xi^\m,
\qquad
g_{ij} \to g_{ij} + \chi_{ij}\,, }
$$
g_{ij} = e^{2 \lambda} h_{ij}\ ,
\ \ \ \ \ \ \
 h_{ij}\equiv
 G_{\m\n}(\bar x) \del_i \bar x^{\m} \del_j \bar x^{\n}
. $$
The classical value of the metric may, in general,
differ from the induced metric $h_{ij}$ by an arbitrary
conformal factor $\l$. We fix the 2-d diffeomorphism invariance
by imposing the conformal gauge conditions on the fluctuations
of the metric
\eqn\fixdiff{
\chi_{ij} = \k g_{ij}\ ,
\qquad
{\rm i.e.}
\qquad
g_{ij} \to (1 +\k) g_{ij}
\,.}
The remaining conformal degree of freedom of the metric
should decouple as in flat 10-d space because of the conformal
invariance of type IIB string theory on \adss\ background
\refs{\mt, \mtt}. To check this we treat $g_{ij}$ as an arbitrary
background metric, {\it not}  identifying it (both in the action
and in the path integral measure) with $h_{ij}$.

Let us introduce the tangent-space components of the fluctuation fields
which have the canonical norms
\eqn\fbzet{\eqalign{
\z^a = E^a_\m \xi^\m,
\qquad
\z^p = E^p_\m \xi^\m,
\qquad
G_{\m\n} = E^a_\m E^a_\n + E^p_\m E^p_\n\,,
\cr
\|\z^a\|^2 = \int d^2 \s \sqrt g \, \z^a \z^a,
\qquad
\|\z^p\|^2 = \int d^2 \s \sqrt g \, \z^p \z^p
.}}
$\zeta^a$ and $\zeta^q$ are the fluctuations of the $AdS_5$ and $S^5$
coordinates respectively (the tangent-space 5-d indices $a,b=0,...,4$
and $p,q=1,...,5$ are raised by the flat 5-d metrics).
We then get the following action for the quadratic fluctuations
(we absorb $1\over 2\pi \a'$ by rescaling the quantum fields)
\eqn\ctoq{
S_{2\rm B} = { 1 \ov 2} \int d^2 \s \sqrt {g}
\left( g^{ij} D_i \z^a D_j \z^a
+ X_{ab} \z^a \z^b
+ g^{ij} D_i \z^q D_j \z^q
+ X_{pq} \z^p \z^q \right),
}
\eqn\yyutrq{
X_{ab} = - g^{ij}\eta_i^{\,c} \eta_j^{\,d} R_{acbd}\,,
\qquad
X_{pq} = - g^{ij}\eta_i^{\,r} \eta_j^{\,s} R_{prqs}\,.
}
Here
\eqn\etttt{
\eta_i^{\,a} \equiv \del_i \bar x^\m E^a_\m\ ,\ \ \ \ \
\ \ \ \ \eta_i^{\,p} \equiv
\del_i \bar x^\m E^p_\m\ ,
}
are the projection of the $AdS_5$ and $S^5$ vielbeins on the
world sheet. $D_i$ is the covariant derivative containing
the projection of the target space spin connection,
\eqn\strcovd{
D_i \z^a = \del_i \z^a + w^{ab}_i\z^b\,,\ \ \ \
\qquad
w^{ab}_i= \del_i \bar x^\m \Omega^{ab}_\m\,
,}
where $\Omega_\mu^{ab}$ is the spin connection of $AdS_5$,
and similarly for $S^5$. For example, the $AdS_5$ part of
the connection is $\Omega^{m4}_\m = - w^{-1} \d^m_\m$,
where $4$ stands for the radial direction and $m=0,1,2,3$.

In general, there will be two types of divergences --
depending on the background $\bar x$ field
$O(\del \bar x \del \bar x)$ (i.e. renormalization of the
target space metric) and proportional to the curvature $\R$
of the fiducial 2-d metric $g_{ij}$ (i.e. renormalization
of the dilaton).

To check the conformal invariance we
need to use the fact that for \adss\
\eqn\forads{
R_{acbd}=-\delta_{ab}\delta_{cd}+\delta_{ad}
\delta_{cb}\ ,
\ \ \ \ \ \
R_{prqs}=\delta_{pq}\delta_{rs}-\delta_{ps}\delta_{rq}\ .
}
Then
\eqn\rewqq{
X^{ab} = g^{ij} \eta_i^{\,c} \eta_j^{\,c} \delta^{ab}
- g^{ij} \eta_i^{\,a} \eta_j^{\,b},
\qquad
X^{pq} = - g^{ij} \eta_i^{\,r} \eta_j^{\,r} \delta^{pq}
+ g^{ij} \eta_i^{\,p} \eta_j^{\,q}\ .
}
The $\bar x$-dependent logarithmic UV divergences coming
from \ctoq\ are proportional to
\eqn\treq{
 \tr\ X\
= 4 g^{ij} \left( \eta_i^{\,a} \eta_j^{\,a}
- \eta_i^{\,p} \eta_j^{\,p}\right)\ .
}
This gives the Ricci tensor dependent of the conformal
invariance equation,
$R_{\hat a\hat b} - { 1 \ov 4 \cdot 4!}
F_{\hat a....} F_{\hat
b....}=0$.
The 5-form dependent part will come from the fermionic
contribution.

$\k$-symmetry transformations in curved \adss\ space \mt\
which leave \frmo\ invariant have a form similar to flat-space
transformations \kap\
\eqn\kapc{
\delta_\k \t^I = \td \r^\dagger_i \k^{i I} + ...\ , }
where
\eqn\ononu{
\frac{1}{\sqrt{g}}\epsilon^{ij}\kappa_j^1=-\kappa^{i1},
\ \ \ \ \ \
\frac{1}{\sqrt{g}}\epsilon^{ij}\kappa_j^2=\kappa^{i2}\ ,
}
and (cf. \nott)
\eqn\nonnn{
\td \r^\dagger_i = \left(\G_a E^a_\m - i\G_p E^p_\m\right)
 \del_i x^\m\ .
}
 Fixing the $\k$-symmetry gauge
\eqn\kaga{ \theta^1=\theta^2=\theta\ ,
}
the quadratic part of the fermionic action \frmo--\nott\ is
found to be
\eqn\feracq{
S_{2\rm F} =
2i \int d^2 \s
\left(\sqrt {g} g^{ij} \bar\theta\r_i\DD_j\theta
-{i\over2}\epsilon^{kj}\bar\theta\r_k\tilde\r_j\theta \right) \ .
}
We shall
first be interested in $\del \bar x \del\bar x$ divergences
so that it should not be necessary to distinguish between
$g_{ij}$ and the induced metric, therefore
\eqn\sotaht{
 \r_{(i} \r_{j)} = g_{ij} =
 G_{\m\n} \del_i \bar x^\m \del _j \bar x^\n
= \eta_i^{\,a}\eta_j^{\,a}+ \eta_i^{\,p}\eta_j^{\,p} \ .
}
In general, given the operator
$ {\cal O}=i \r^k \DD_k + M$, the logarithmic divergence in
$\ha \ln \det({\cal O} {\cal O}^\dagger)$
is proportional to\foot{Starting with
$( i\r^k \del_k + M) (i\r^n \del_n + M^\dagger)$
one is to note that in the general
case of $ \r^k M^\dagger + M \r^k \not=0$ one is to
introduce an additional connection to put the resulting operator in the
standard form
$- D^2 + X$.}
\eqn\prwe{
 - { 1 \ov 4} g^{ij} \tr ( \r_i M \r_j M + \r_i M^\dagger \r_j
 M^\dagger) \ , }
where in the present case
$M= {1\over2}{\epsilon^{kj}\over\sqrt g}\r_k\tilde\r_j, \
M^\dagger = -{1\over2}{\epsilon^{kj}\over\sqrt g}
\tilde\r^\dagger_k\r_j$,\ \
$\r_i = \r_i^\dagger$.
After some algebra\foot{If the background does not depend on $S^5$,
then $\r_i=\tilde\r_i$ and
the calculation is trivial. }
one finds that \prwe\ reduces to
\eqn\gammasq{
-4\left[\det\left(\eta_i^{\,a}\eta_j^{\,a}\right)
-\det\left(\eta_i^{\,p}\eta_j^{\,p}\right)\right]
=4
g^{ij}\left(\eta_i^{\,a}\eta_j^{\,a}-\eta_i^{\,p}\eta_j^{\,p}\right)
\,.}
This combination determines the fermionic contribution
to the $\bar x$-dependent logarithmic divergence, and it
exactly {\it cancels} the bosonic contribution \treq.

As usual, fixing conformal gauge produces
bosonic ghosts which are 2-d vectors
\eqn\ghostqq{
S_{\rm gh} = { 1 \ov 2} \int d^2 \s \sqrt {g} g^{ij}
\left(g^{kl} \nabla_k\epsilon_i \nabla_l \epsilon_j
-\ha\R \epsilon_i\epsilon_j \right),
}
where $\R$ is the scalar curvature of the 2-d metric $g_{ij}$.

The quadratic and
linear divergences cancel between bosons and fermions
because of the matching of the number of degrees of freedom.
The coefficients of the logarithmic divergence
for the above system of fields (the Seeley coefficients of the
corresponding second order Laplace operators)
are
\eqn\seelbq{
b_{2\rm B}=10\times{\R\over6}-\tr X\,,
\qquad
b_{2\rm gh}=-2\times{\R\over6}-\R\,,
\qquad }
$$
b_{2 \rm F}
=8\times{\R\over 3}+\tr X\,.
$$
Here we took into account that
since the kinetic part of the fermionic operator
depends on the background 2-d metric through
$\sqrt g g^{ij}$,
 the $\R$-dependent part
of its divergence and conformal anomaly coefficient
is four times greater than for a 2-d Majorana fermion,
just like
in the flat GS string case \wrong\
(this difference may be attributed to the contribution of
the Jacobian of a local rotation that transforms
$\r_i$ into 2-d Dirac matrices contracted with zweibein,
 see Appendix C).

The total divergence coefficient is then
\eqn\seeltq{
b_2^{\rm total}
=\left(10-2+8\times 2\right)\times{\R\over6}-\R
=3\R\,.
}
As was mentioned above, at 1-loop order the argument for the cancellation 
of logarithmic divergences is identical to the argument 
in the case of the flat GS string, where \seeltq\ is also valid. 
Integrating over the scalar curvature on a closed surface will give the 
Euler character $\int d^2\s\sqrt g\,3\R=12\pi\chi$. The same is true on a 
surface with boundary where, as was shown in \alv, all the factors of $\R$ 
are accompanied by the appropriate boundary term. 
Now one should remember that the cutoff dependent 
factors in the conformal Killing vector and/or Teichm\"uller measure 
exactly cancels this divergence, so the final result is $(D-10)\chi$, 
namely zero. This is, of course, consistent with the cancellation
of the total $\bar x$-independent conformal anomaly, or the central 
charge, ensuring that the dilaton equation is satisfied. Note that this 
is just a consequence of working in the critical dimension.

Thus, we have
confirmed that the theory is conformal at one loop.
As was argued in \refs{\mt,\mtt},
this should be true to all orders in $\a'$ expansion
(for example, the first non-trivial correction
to the central charge vanishes because
the Ricci scalar of the target space metric is zero,
$R_{tot} = R_{AdS_5} + R_{S^5} =0$, etc.).

\subsec{\bf Nambu-type action in the static gauge}
Alternatively, one may start with the corresponding Nambu-Goto
form of the GS action (with no independent 2-d metric).
This action is highly non-linear, but in the quadratic
approximation it is straightforward to determine the 2-d
operators of small fluctuations of a string in
curved background. Here it is natural
to choose the static gauge to fix the diffeomorphisms,
i.e. to identify the world sheet coordinates
with the two target space coordinates and demand that there
are no fluctuations in those directions. The ghost
determinant is then ``local", i.e. is a determinant of an
operator of multiplication by a function (which needs a
regularization and may still produce non-trivial contribution
to partition function).

Following the standard procedure (for a careful treatment see,
e.g., \moor), fixing the static gauge
$\delta x^k=0$, \ $k=0,1$,\ produces the ghost determinant
\eqn\ghost{
\Delta_{\rm gh}^{-1}=
\int [d \ep] \exp \left( - \ha \| \delta_\ep x^k\|^2\right)\,
. }
The path integral over the 2-d diffeomorphism parameters
$\ep^i$ is defined using the norm
\eqn\nori{
\|\ep\|^2 = \int d^2 \s \sqrt h h_{ij} \ep^i \ep^j
.}
Here $h_{ij}$ is the induced
metric \smlfl
\eqn\indmet{
h_{ij}=G_{\m\n}(\bar x)\del_i\bar x^\m \del_j \bar x^\n
=\eta_i^{\,\hat a}\eta_j^{\,\hat a}
\,.}
Explicit evaluation of \ghost\ gives
\eqn\diif{
\|\delta_\ep x^k\|^2 =
\int d^2 \s \sqrt h G_{kl} (\bar x) ( \ep^i \del_i \bar x^k)
( \ep^j \del_j \bar x^l)
=
\int d^2 \s \sqrt h h^\parallel_{ij} \ep^i \ep^j \,,
}
$$ h^\parallel_{ij}\equiv G_{kl}(\bar x)\del_i\bar x^k \del_j \bar x^l\ ,
$$
so that
\eqn\ghod{
\Delta_{\rm gh} = [\det ( h^\parallel_{ik} h^{kj} )]^{1/2}
\ . }
As we will see on the examples discussed below, the most natural
regularization of this ``local" determinant is by changing
the normalization of some of the fluctuating fields.

Another possible gauge is to remove the vielbein components
of the longitudinal fluctuations. It is easy to see, by the
same calculation, that the ghost determinant is equal to 1 in
that gauge.

The resulting bosonic action is a modification of
\ctoq
\eqn\ctoqst{
S = { 1 \ov 2} \int d^2 \s \sqrt {h}
\left( h^{ij} D_i \z^{\bar a} D_j \z^{\bar a}
+ \bar X_{\bar a\bar b} \z^{\bar a} \z^{\bar b}
\right)\,,
}
where $\zeta^{\bar a}$ are the fields representing the transverse
fluctuations. $\bar X$ is not the same as $X$ (in the simple
examples discussed below $\tr\bar X=\tr X+\R$).
The fermions are treated in the
same way as before, so squaring the fermionic operator
gives a mass term whose trace is equal to $\tr X$.

The non-trivial $O(\del \bar x \del \bar x)$
part of divergences cancels
again, while the remaining $\int \R(h) $ part
(which, in the presence of a boundary should
be accompanied by an
appropriate boundary term to give the Euler number)
should be canceled in $D=10$
by appropriate measure factor contributions,
as happens in conformal gauge
\alv.

It should be stressed that, while the result of a semiclassical
computation in the Nambu action case should be equivalent to
the one in the Polyakov action case \ft, a careful definition
of the path integration measure is non-trivial in the Nambu case.
For that reason we prefer to use the Polyakov definition of the
string partition function which is well-defined. It should be
clear that the problem of cancellation of $\int \R$ divergences
is exactly the same as in the case of the Nambu action in flat
space, and thus has nothing to do with peculiarities of the
\adss\ background. This resolves the puzzle of the apparent
non-cancellation of logarithmic divergences that was encountered
in \fgt, which is revealed to be an artifact of the use
of the Nambu-type formulation without including the additional
measure contributions to the divergences and ignoring the
subtleties of the divergences/conformal anomaly
cancellation in the flat-space GS superstring theory.

We shall see that in the cases of interest the kinetic operator
of GS fermions takes (after a local rotation) the form of the
2-d Dirac operator in the curved 2-d geometry defined by the
induced metric. In view of the above discussion, the fact that
when one directly evaluates the divergences one seems to
find that the $\R$-terms do not cancel is an artifact
of not distinguishing between the generic and the induced metrics.
These topological divergences are, in any case, irrelevant
for the evaluation of the non-trivial part of the partition
function which determines the correction to the Wilson loop
expectation value, or to the $1/L$ potential.
The issue of divergences may be avoided altogether,
by normalizing, as we suggest below, the partition function to
its value for some standard background.

Once this issue has been clarified, one should be able to use the
static gauge expressions for the small-fluctuation determinants,
as they sometimes turn out to be simpler than the analogous
expressions in the Polyakov formulation in conformal gauge.
In what follows we shall not distinguish between
the generic fiducial metric $g_{ij}$ and the induced metric
$h_{ij}$ (using always the notation $g_{ij}$ for the 2-d metric).

\subsec{\bf Relating quadratic GS fermion term
to 2-d Dirac fermion action }

Before turning to specific examples let us make some general
comments on how one can put the quadratic fermionic term
\feracq\ in the GS action in curved target space background
into the standard kinetic term for a set of 2-d fermions defined
on a curved 2-d space. The main idea is to apply a local
target space Lorentz rotation to GS spinor $\t$, as discussed
previously in the case of the heterotic string in flat
\refs{\sedr,\leu,\wig}  and curved \refs{\kara,\lech }
spaces. We shall concentrate on the derivative term in the
fermionic action \feracq. It should be noted that the presence
of the second ``mass" term in \feracq\ which originates from
the coupling to RR background and which is absent in the
heterotic case will not allow to compute the resulting
2-d fermion determinant in a closed form using the standard
anomaly \pow\ arguments. Ignoring the distinction between
$g_{ij}$ and the induced metric \indmet\ we can write the
derivative term in \feracq\ as
\eqn\feacq{
S^{(deriv.)}_{2\rm F} =
2i \int d^2 \s
\ \sqrt {g} g^{ij} \bar\theta\r_i\DD_j\theta
= { i }
\int d^2 \s
\ \sqrt {g} g^{ij} (
 \bar\theta\r_i\del_j\theta
 - \del_j\bar\theta\r_i \theta)
 \ . }
Let us introduce the tangent $t^\m_\a$ ($\m=0,1,...,9$, \
$\a=0,1$) and normal $n^\m_s$ ($s=1,...,8$)
vectors to the world surface which form orthonormal 10-d basis
($g_{ij} = e^\a_i e^\b_j \eta_{\a\b}$)
\eqn\veco{
t^\m_\a = e^i_\a \del_i \bar x^\m \ , \ \ \ \ \
(t_\a, t_\b) =\eta_{\a\b} \ , \ \ \
(t_\a, n_s)=0 \ , \ \ \ \
(n_s, n_u)= \delta_{su} \ , }
where $(a,b) = G_{\m\n} a^\m b^\n$.
Then one can make a local $SO(1,9)$ rotation of this basis
which transforms the set of $\s$-dependent
10-d Dirac matrices (see \trw) into the 10 constant
Dirac matrices
\eqn\sett{
\r_\a (\s) = e^i_\a \r_i = S(\s) \G_\a S\inv (\s) \ ,
\ \ \ \
 \r_s (\s) = n^\m_s E_\m^{\hat a} \Gamma_{\hat a}
 = S(\s) \G_s S\inv (\s) \ .
}
One may further choose a representation in which
$\G_\a= \tau_\a \times I_8$, where $\tau_\a$ are 2-d Dirac
matrices.  Depending on the specific embedding
and particular curved target space metric, one may then be
able to write the action \feacq\ as the action for 2-d Dirac
fermions coupled to curved induced 2-d metric and interacting
with some gauge fields (coming from $S\inv dS$).

Simple examples when this happens will be discussed below.
We shall consider embeddings of the string world sheet into
the $AdS_3$ part of the $AdS_5$ space, so there will be
only {\it one } normal direction and the extra normal bundle
2-d gauge connection will be absent (cf. \refs{\sedr,\leu,\kara}).
In this 3-dimensional embedding case with non-chiral 2-d fermions
the Jacobian associated with the local Lorentz rotation will be
trivial (see also Appendix C).

\newsec{One-loop approximation near the straight string
configuration: Supersymmetric field theory on $AdS_2$}

The simplest classical solution for string in \adss\
is a straight string with the world surface
spanned by the radial direction of $AdS_5$ and time.
The Euclidean solution and the corresponding induced metric are
\eqn\strlin{
\tau=x^0\,,
\qquad
\sigma=x^4=w\,,
\qquad
ds^2 = { 1 \ov \s^2} (d\tau^2 + d\s^2)\,
.}
The induced metric on the world sheet is that of $AdS_2$, \
with constant negative curvature $\R=-2$ \ (the radius of
$AdS_5$ is $R=1$).

This solution represents a single straight Wilson line
running along the Euclidean time direction. This is a BPS
object in string theory, it
corresponds to a static fundamental string stretched between
a single D3-brane and $N$ coinciding D3-branes. Therefore,
one would expect that the partition function be
equal to 1. The properly defined (subtracted) classical
string action evaluated on \strlin\ indeed vanishes, and we
shall evaluate the 1-loop correction to the partition function.

As we shall show, the corresponding 1-loop correction to the
vacuum energy defined with respect to a certain time-like
Killing vector vanishes. Relating the vacuum energy
to the partition function using a conformal rescaling argument
(and the fact that the total conformal anomaly is zero) 
we conclude that $Z=1$. It should be mentioned that
while the (properly defined) vacuum energy of a supersymmetric
field theory in $AdS$ space should vanish, this does {\it not}
automatically imply (in contrast to what happens in flat space)
that the partition function of such theory should be equal to 1
(cf. \refs{\barf,\cahi}). In the present  case this
 happens only with the
inclusion of the appropriate ghosts and longitudinal modes.
The calculation of the partition function is rather subtle,
and depends on a regularization prescription.
Let us note also that 
 the point of view of physical
applications, the precise value of $Z$ (which is simply a constant)
is not actually important, and one may normalize with respect to
it in computing $Z$ for more general string configurations.

Apart from being the simplest example, there are other reasons
why the analysis of the straight string case is of interest.
Any smooth Wilson loop looks in the UV region like a straight line.
In the present set-up this translates into the behavior of
the minimal surface near the boundary of $AdS_5$ space. In the
general case one will have to calculate the partition function
for a complicated two dimensional field theory. But
asymptotically the minimal surface will approach $AdS_2$, and the
small fluctuation operators (in particular, the asymptotic
values of the masses of the fluctuation fields) will also be
the same as for a straight string. Many subtleties related to
divergences and asymptotic boundary conditions are already present
in this example, and they can be automatically avoided in more
general cases by normalizing with respect to the partition
function of the straight string.

\subsec{\bf The action and multiplet structure}

The bosonic part of the action for small fluctuations
in conformal gauge is \ctoq\
\eqn\pto{
S_{2\rm B} =
\ha \int d^2 \s \sqrt {g}
\left( g^{ij} D_i \z^a D_j \z^a + X_{ab} \z^a \z^b
+ g^{ij} D_i \z^p D_j \z^p\right)
,}
where in the present case
\eqn\beinproj{
\eta_0^{\,a} = \partial_0\bar x^\mu E_\mu^a = w^{-1}(1, 0, 0,0,0)\,,
\qquad
\eta_1^{\,a} = \partial_1\bar x^\mu E_\mu^a = w^{-1}(0, 0, 0,0, 1)
\,.
}
$\eta_i^{\,a}$ with $a=0,4$ is thus just a
vielbein of the induced 2-d metric $g_{ij}$,
\eqn\vier{
\eta_i^{\,a} = e_i^{\,\alpha}\ , \ \ \ \ \ \ \ \
a=0,4\ , \ \ \ \a =0,1 \ , }
so that
\eqn\strx{
X_{ab} = {\rm diag} (1,2,2,2,1)\,.
}
The only nonzero connection in $D_i$ \strcovd\ is $w_0^{04}
=-w^{-1}$, i.e.
\eqn\strcovds{\eqalign{
D_0 \z^0 = \del_0 \z^0 - w^{-1}\z^4,
\qquad\ \ \ \
D_0 \z^4 = \del_0 \z^4 + w^{-1}\z^0.
}}
The natural norms for the fields are
\eqn\strmeas{
\|\z^{\hat a}\|^2
= \int d^2 \s \sqrt g \z^{\hat a} \z^{\hat a}
= \int d \tau d \s { 1 \ov \s^2} \z^{\hat a} \z^{\hat a}
.}
The ghost action is the same as in \ghostqq, i.e.
\eqn\asdeeww{
\ha\int d^2 \s
\sqrt g (\nabla^i \ep^\alpha \nabla_i \ep^\alpha
- \ha \R \ep^\alpha \ep^\alpha)\,,}
where we defined $\epsilon^\alpha=e^\alpha_i\epsilon^i$
with flat 2-d tangent space indices, and $\nabla_i$ includes
the world sheet Lorentz connection.

Because of the direct embedding of the world sheet into the
target space there are some extra simplifications. The projection
of the target space connection on the world sheet $w_i^{ab}$ is
the same as the spin connection of the induced metric appearing
in $\nabla_i$. In addition, $-\ha \R=1$. Therefore, the action of the
ghosts is identical to the action of the longitudinal modes
$\zeta^0$, $\zeta^4$, but the boundary conditions are different
\alv.

Before $\k$-symmetry gauge fixing the fermionic Lagrangian \frmo\
is (here we use Minkowski notation)
\eqn\iasd{
L_{2\rm F}
=
-i\left(\sqrt{g}g^{ij}\delta^{IJ}-\epsilon^{ij}s^{IJ}\right)
\bar\theta^I\rho_i D_j\theta^J\ ,
}
where in the present case
\eqn\aisd{\eqalign{
\rho_i
&=
e_i^{\a}\r_{\a}
=\eta_i^{\,a}\Gamma_a
=\left\{\matrix{
w^{-1}\Gamma_0 & \hbox{for} & i=0\,,\cr
w^{-1}\Gamma_4 & \hbox{for} & i=1\,,
}\right.
\cr
D_i\theta^J
&=\hat\nabla_i\theta^J-{i\over2}\epsilon^{JK}\rho_i\theta^K\,,
\cr
\hat\nabla_0
&=
\partial_0-{1\over 2w}\Gamma_{04}\,,
\qquad
\hat\nabla_1=\partial_1\,.
}}
We choose again the gauge $\theta^1=\theta^2=\theta$.
Then
\eqn\fuum{
L_{2\rm F} = -2i \sqrt{g} \big( \bar \theta \rho^i {\hat \nabla}_i \theta
+ i \bar \theta \r_3 \theta \big),
\qquad
\r_3
 \equiv \ha \ep^{\a\b} \r_\a\r_\b
 = \Gamma_{04}\,
.}
Here we introduced the notation $\r_\a=(\G_0,\G_4)$
($\r_\a$ may be identified with 2-d Dirac matrices times $I_8$).
Thus the quadratic fermionic part of GS action has exactly the
same form as the action for 2-d fermions in curved 2-d space.

Assuming the standard $\int d^2 \s\sqrt{g} \bar\theta\theta$
normalization, the corresponding Dirac operator is
\eqn\asdwed{
D_F = i \r^{ i} \hat\nabla_i - \r_3
= i w ( - \G_0 \del_0 + \G_4 \del_1)
 - \ha i \G_4 - \G_0 \G_4
\,,}
where the third term came from $D_0$. The spectral problem is thus
\eqn\asdqqe{
[ i w ( - \G_0 \del_0 + \G_4 \del_1)
 - \ha i \G_4 - \G_0 \G_4] \theta = \l \theta
\,.}
Directly squaring this operator we get
\eqn\asdwee{
\bigg(
w^2 [(\del_0 - \ha w^{-1}
 \G_0\G_4)^2 - \del^2_1 ]
 + \ha \bigg) \theta
=
(- \hat \nabla^2 + { 1 \ov 4} \R + 1 ) \theta = \l^2 \theta
\,.}

Ignoring the ghosts and longitudinal modes, we are left with a 2-d
field theory on $AdS_2$ containing five massless scalars, three
scalars with mass squared 2, and eight fermions with mass squared 1.
Field theories on $AdS_2$ were studied in the past
(see \refs{\sakone,\sakpot,\saktan,\ina,\bul,\barf,\iopot}).
The fields in the $\cN=1$ scalar supermultiplet in $AdS_2$
may have the following bosonic and fermionic masses
\refs{\ivan,\barf,\sakpot}:
\eqn\styy{
m^2_B = \m^2 - \m\,,\ \ \
\qquad
m_F = \m
\,,}
where $\m$ is a free parameter. In the case at hand we have 5
``massless'' multiplets with $\mu=1$ ($m^2_B = 0$, $m_F = 1$)
and 3 ``massive" multiplets with $\mu=-1$ ($m^2_B = 2$, $m_F = -1$).
It is possible to combine a $\mu=1$ and a
$\mu=-1$ multiplet into an $\cN=2$ multiplet, the dimensional
reduction of the 4-d chiral multiplet to 2 dimensions.
Two $\mu=1$ multiplets also form an $\cN=2$ multiplet which
is the dimensional reduction of the 4-d vector multiplet.
Three chiral and one vector multiplets in $D=4$ make
one $\cN=4$ vector in four dimensions, so we conclude that
the 8 scalars and 8 fermions that we have found should
form one $\cN=8$ multiplet in two dimensions
(see, e.g., \sprad\ for related discussion).\foot{The mass term
in the action \fuum, contains the matrix $\r_3=\G_0\G_4$ which has
half of its eigenvalues $1$ and half $-1$, i.e. there are
actually 4 fermions with $m_F=-1$ and 4 with $m_F=1$, not $3+5$.
But the sign choice in \styy\ $m_B^2=\m^2-\mu$ rather than
$m_B^2=\m^2+\m$ is for $\cN=1$ supersymmetry. For extended
supersymmetry both signs are possible, so the bosons can be split
into 3 and 5 while the fermions are split to 4 and 4.}

We finally obtain the following partition function
with the scalar and spinor Laplace operators defined
with respect to the Euclidean $AdS_2$ metric with radius 1 \
($\R=-2$)
\eqn\arti{
Z_{B+F}=
{\det^{8/2}\left( - \hat \nabla^2 + \four \R + 1\right)
\over \det^{3/2}\left( - \nabla^2 + 2 \right)
\det^{5/2} \left( - \nabla^2 \right)}
\,.}
It seems reasonable to impose, as is usually done in discussions of
supersymmetric theories in $AdS_n$ backgrounds \refs{\avi,\breit},
proper boundary conditions consistent with supersymmetry. Those
imply that the resulting spectra of Laplace operators are discrete
in spatial direction (and not continuous as one would normally
expect to find in a non-compact hyperbolic space).

A direct calculation of the partition function would involve
solving the spectral problems \asdwee\ and
\eqn\stsp{
\left(-\partial_0^2-\partial_1^2+{m^2\over \sigma^2}\right)\zeta
={\lambda\over \sigma^2}\zeta
\,,}
with $m^2=0, 2$.
The solutions to the bosonic problem which vanish at $\sigma=0$ are
\eqn\bessd{
\zeta(\tau,\sigma)=e^{ip\tau}\sqrt{\sigma}K_{i\nu}(p\sigma)\,,
\qquad
\ \nu^2=\lambda-m^2-\fourth\,,\qquad 0\leq\nu<\infty\,
, }
where $K_{i\nu}$ are modified Bessel functions.

A calculation of the partition function based on this spectrum
is presented in Appendix~B. This direct approach suffers from
regularization problems, it also does not capture the symmetries
of the problem, like supersymmetry. Here we use a different
method to evaluate it.

\subsec{\bf Vacuum energy}


Instead of calculating the partition function directly, we
could start with the vacuum energy. It is given by the determinant
of the operator scaled to remove the factor of $g^{00}$ from
in front of $\partial_t^2$ (see \cahi). Then we use the
conformal anomaly, as discussed in Appendix A, to derive the
partition function.

Let us change the world sheet coordinates so that the $AdS_2$ metric
is
\eqn\adstnm{
ds^2={1\over\cos^2\rho}\left(dt^2+d\rho^2\right)\,,
\ \ \ \ \ \ \ \ \rho\in[-{\pi\over2},{\pi\over2}]\ .
}
The spectra of the Hamiltonians conjugate to this time
variable were calculated in \saktan:\foot{From
the group-theoretical point of view, the unitary irreducible
representations of the $AdS_2$ superalgebra contain:
for $\m > \ha$ a scalar field with $\om^{(B)}_n = n + \m $ and
a fermion field with $\om^{(F)}_n = n + \m + \ha, $
and for $\m < -\ha$ --
a scalar field with $\om^{(B)}_n = n + 2|\m| $ and
a fermion field with $\om^{(F)}_n = n + |\m| + \ha. $ }
\eqn\sutr{
\om^{(F)}_n (\m) = n + |\m| + \ha\ ,
\qquad\ \ \
\om^{(F)}_n (\pm 1)=n + { 3 \ov 2}\ ,
}
$$
\om^{(B)}_n (\m) = n + h(\m)\ ,
\qquad\
h(\m) = \ha( 1 + \sqrt{ 1 + 4m^2_B})\ ,
\qquad
h(-1) =2,
\ \ \
h(1) =1\ . $$
Summing over all the modes we get, as in \sakpot,
the 1-loop vacuum energy of this effective 2-d field theory
\eqn\rqtes{
E= \ha \sum_{n=0}^\infty\left(
3\left[\om^{(B)}_n (-1) -\om^{(F)}_n (-1)\right]
+ 5 \left[\om^{(B)}_n (1) -\om^{(F)}_n (1)\right] \right).
}
As was extensively discussed in the literature, the
{\it properly defined} vacuum energy should {\it vanish}
in the $AdS$ case as it does in flat space
\refs{\sakone,\saktan,\barf,\gibbo,\burg} (even though divergences
may not cancel out, unless there is a lot of supersymmetry \ald).
However, the direct computation of the sum of the mode energies
using $\zeta$-function regularization may lead to a non-zero
result because the $\zeta$-function
regularization may not, in general, preserve supersymmetry.

Using the standard relations
\eqn\stlzet{
\zeta(s,x) \equiv \sum_{n=0}^\infty (n+ x)^{-s},
\qquad
\zeta(-1,x) = - \ha \left(x^2-x + {1 \ov 6}\right)
\,,}
we find for a boson ($m_B^2=\m^2-\m$) 
\eqn\zetab{
E_B
=\ha\sum_{n=0}^\infty [n+h(\m)]
=\ha\z(-1,h(\m))
=-\fourth\left(m_B^2+{1\over6}\right)\,,
}
and for a fermion ($m_F =\m $) 
\eqn\zetab{
E_F
=-\ha\sum_{n=0}^\infty\left(n+|\m|+\ha\right)
=-\ha\z\left(-1,|\m|+\ha\right)
=\fourth\left(m_F^2-{1\over12}\right)\,.
}

In our case we find\foot{Note that
if we set the mass terms to zero by taking $\mu=0$ in \styy,
we get 8 massless scalars and 8 massless fermions in $AdS_2$ and
their vacuum energies do not cancel -- we get
$\ha\times8[\zeta(-1,1)-\zeta(-1, \ha )]=4\times(-{1\ov 8})$.
}
\eqn\vacuu{\eqalign{
E
=-\fourth\left[3\times\left(2+{1\over6}\right)
+5\times {1\over6}
-8\times\left(1-{1\over12}\right)\right]
=0\,
.}}
The ratio 3:5 of the numbers of the two multiplets
is just what is needed for the cancellation.

The fact that $E$, defined by $\zeta$-function regularization,
vanishes may be a consequence of the extended $\cN=8$
supersymmetry mentioned above. Indeed, while as was originally
suggested \sakone\ for $AdS_4$ and confirmed also in
\refs{\barf,\gibbo,\burg}, the $\zeta$-function regularization
may break supersymmetry and thus may lead to $E\not=0$,
this does not actually happen in the case of $N\geq 5$, $D=4$
gauged supergravities \ald. The present $D=2$ case is thus
analogous to those $D=4$ cases with large amounts of
supersymmetry.\foot{The $E=0$ property of $N\geq 5$, $D=4$
supergravities might be related to the cancellation \fini\
of the logarithmic gauge coupling renormalization in these
theories (note, in particular, that as was discussed in \mye\
in the case of the flat space, the vacuum energy as defined by
the partition function is the same as the sum of zero-point
energies provided $\sum \zeta(0)=0$, i.e. if there are no UV
infinities). It may seem that the analogy between our $D=2$
case and the $D=4$ cases is not quite complete since here,
in fact, the naive calculation of the coefficient of the
logarithmic divergence in terms of the sum of $\zeta$-functions
gives a nonzero answer (using $\zeta(0,x)=\ha -x$, we find the
total coefficient to be 1). Note, however, that the types of
divergences which cancel in the $D=4$ cases and do not cancel
in the $D=2$ case are actually quite different, i.e. the direct
comparison is not possible.}

\subsec{\bf Partition function}

The vacuum energy calculated in the previous subsection as the
sum over oscillator modes corresponds to the determinants 
of  the following (mass $m$)  bosonic and fermionic 
spectral
problems
\eqn\venspp{\eqalign{
\left(-\del_0^2-\del_1^2+{m^2\over\cos^2\rho}\right)\z
=\l_B\z\,,
\cr
\left(-\hat\nabla_0^2-\hat\nabla_1^2+\ha{1\over\cos^2\r}\right)\t
=\l^2_F\theta\,,
}}
where in the fermionic operator we assume
that the covariant derivatives are contracted using flat metric.
These are related to the spectral problems \asdwee\ and \stsp\
(apart from the coordinate change) by a rescaling of the right hand
side by $\cos^2\r$. As  was mentioned above,
 in curved (e.g., static conformally flat)
space the logarithm of the 
partition function is, in general, 
 different from the vacuum energy
defined as a sum over eigen-modes because the time derivative
part of the relevant elliptic operators is rescaled by $g^{00}$.
The  determinants of the two operators 
which differ by such a rescaling 
are related to each other
by a  conformal anomaly type equation 
as discussed in Appendix A. 
The extra contribution from a mass $m$ boson is
\eqn\exbm{
-\log\det\D_1=-\log\det\D_M
+{1\over4\pi}\int d^2\s\sqrt g
\left(m^2\ln M+{1\ov12}\del^i\ln M\del_i\ln M\right)\,,
}
where $M=\cos^2\r=1/\sqrt g$. The two terms in the parentheses 
differ only by a total derivative, but we choose to write it this 
way to eliminate the boundary terms. Each fermion contributes
\eqn\exef{
\log\det\D_1=\log\det\D_{{\cal K}^2}
-{1\over4\pi}\int d^2\s\sqrt g
\left(2\ln {\cal K}
-{2\ov 3}\del^i\ln {\cal K}\,\del_i\ln {\cal K}\right)\,,
}
where ${\cal K}=\cos\r=\sqrt M$. 
Summed together the  `transverse' scalars and fermions contribute
\eqn\exef{
\log Z_{B+F}
=-{1\over4\pi}\int d^2\s\sqrt g
\left(\ln M - \del^i\ln M\del_i\ln M\right)\,.
}
To this we should add  the contribution of the ghosts and
longitudinal modes. For the ghosts one gets the standard 
Liouville action
\eqn\exgh{
\log\det\D_1^{\rm gh}=\log\det\D_M^{\rm gh}
-{26\ov12}\times {1\over4\pi}\int d^2\s\sqrt g\,
\del^i\ln M\del_i\ln M\,.
}
The longitudinal modes have the same action as the ghosts, but 
different boundary conditions giving the conformal anomaly
\eqn\exlm{
-\log\det\D_1^L=-\log\det\D_M^L
+2\times{1\over4\pi}\int d^2\s\sqrt g
\left(\ln M+{1\ov12}\del^i\ln M\del_i\ln M\right)\,,
}
so that 
\eqn\exef{
\log Z_{L+{\rm gh}}
={1\over4\pi}\int d^2\s\sqrt g
\left(\ln M -  \del^i\ln M\del_i\ln M\right)\,.
}
Putting it all together we find that the partition function is
identically equal to  one
\eqn\totpf{
Z_{\rm total}
={\det^{1/2}(-\D^{\rm gh}_{ij}+\d_{ij})
\det^{8/2}(-\hat\nabla^2+{1\over4}\R+1)
\over
\det^{1/2}(-\D_{ij}+\d_{ij})
\det^{3/2}(-\nabla^2+2)
\det^{5/2}(-\nabla^2)}
=1
\,.}
This result is a consequence of 
  the fact  that the vacuum energy vanishes, 
and also  of  the cancellation of the sum of 
 conformal anomalies for ten bosons with
total mass terms 8, eight fermions with 4 times the
standard 2-d  fermion conformal
anomaly and total mass 8, and the conformal gauge  ghosts.

Note that this is not identical to the conformal
 anomaly calculation of
Section 3.2. Here we did {\it not} distinguish between
the induced and the fiducial metric. An alternative method
of calculating the partition function would be to go back 
and treat the fiducial metric $g_{ij}$ and the induced
metric $h_{ij}$ as independent. Then only the fiducial 
metric should be rescaled, while  the induced metric should 
not. It is most convenient to work with flat metric on the 
strip
\eqn\flonst{
g_{ij}=\delta_{ij}\,,
\qquad
h_{ij}={1\over\cos^2\rho}\delta_{ij}\,.
}
That eliminates the problem of the boundary contributions, 
since the geodesic curvature is zero. This calculation gives 
the same spectral problem as the vacuum energy calculation 
for the bosons and ghosts, but not the fermions and 
longitudinal modes.

\newsec{Circular Wilson loop}

Another case where the classical solution has an explicit
simple form \refs{\dcfm,\dgo} is a circular Wilson loop.
Like the straight string case, this configuration is useful
as it gives a laboratory to investigate many of the issues
that arise in the case of the more general bent string
configuration.

\subsec{\bf Classical solution and quadratic fluctuation action}

The target space metric in polar coordinates is
($s=2,3$)\foot{Because of scale invariance,
the radius of the circle may be set equal to one.}
\eqn\asdccd{
ds^2={1\over w^2}\left(dr^2+r^2d\phi^2+dx^sdx^s+dw^2\right)
+d\Omega_5^2
\,.}
We set $x^0=\phi\in[0,2\pi]$ and
$x^1=r\in[0,1]$.

The classical solution and the induced metric are
\eqn\asdcdd{
w=\sqrt{1-r^2}\,,
\qquad
g_{ij}=\pmatrix{{r^2\over w^2}&0\cr0&{1\over w^4}},
\qquad
\ \
\sqrt{g}g^{ij}=\pmatrix{{1\over rw}&0\cr0&rw}
\,, }
\eqn\asdddc{
\R=-2\ ,
}
i.e. the world sheet metric is again that of $AdS_2$
\dcfm.\foot{
To put the metric in a more standard form we set
$y=w^{-1}$. Then
$ds^2=(y^2-1)^{-1}dy^2+(y^2-1)d\phi^2$,
or in terms of $\tanh\chi=r$,
$ds^2=d\chi^2+\sinh^2\chi d\phi^2$.
}

In the conformal gauge the quadratic part of the bosonic action is
\ctoq, i.e.
\eqn\ctoqc{
S =
\ha \int d^2 \s \sqrt {g}
\left( g^{ij} D_i \z^a D_j \z^a
+ X_{ab} \z^a \z^b
+ g^{ij} D_i \z^q D_j \z^q
\right)
\,,}
where in the present case
$$X^{ab}=2\delta^{ab} - g^{ij}\eta_i^{\,a}\eta_j^{\,b}\,,
\qquad
\eta_0
=\left({r\over w},0,0,0,0\right),
\qquad
\eta_1
=\left(0,{1\over w},0,0,-{r\over w^2}\right)\ . $$
The nonzero components of the spin connection in the
target space are
\eqn\spcots{
\Omega_0^{01}=1\,,
\qquad
\Omega_0^{04}=-{r\over w}\,,
\qquad
\Omega_1^{14}=
\Omega_2^{24}=
\Omega_3^{34}=
-{1\over w}\,,
}
so that all of the covariant derivatives are trivial,
 $D_i=\partial_i$,
except for
$$
D_0\zeta^0=\partial_0\zeta^0+\zeta^1-{r\over w}\zeta^4,
\qquad
D_0\zeta^1=\partial_0\zeta^1-\zeta^0,
\qquad
D_0\zeta^4=\partial_0\zeta^4+{r\over w}\zeta^0,
 $$\eqn\except{\eqalign{
D_1\zeta^1=\partial_1\zeta^1-{1\over w}\zeta^4,\ \ \
\qquad
D_1\zeta^4=\partial_1\zeta^4+{1\over w}\zeta^1. }}
The covariant derivative $\nabla_i$ in the ghost action
\eqn\ghostc{
S =
\ha \int d^2 \s \sqrt {g}
\left(g^{ij} \nabla_i\epsilon^\alpha \nabla_j \epsilon^\alpha
-\ha\R \epsilon^\alpha\epsilon^\alpha \right),
}
includes the world-sheet
spin connection, whose only nonzero component is
$\omega_0^{01}=w^{-1}$. This derivative is not the same as the covariant
derivative
 in \except. However, if we rotate the fields
\eqn\rotate{
\pmatrix{\tilde\zeta^1\cr\tilde\zeta^4}
=
\pmatrix{\cos\alpha & -\sin\alpha \cr \sin\alpha & \hfill\cos\alpha}
\pmatrix{\zeta^1\cr\zeta^4},
\ \ \
\cos \alpha=w\,,
\ \
\sin \alpha=r\,,
\ \
{d\alpha\over dr}={1\over w}\,,
}
the mass matrix becomes diagonal,
$\tilde X_{ab}=\diag\left(1,1,2,2,2\right)$, and the
only nontrivial covariant derivatives are
$D_0\zeta^0=\partial_0\zeta^0+w^{-1}\tilde\zeta^1$ and
$D_0\tilde \zeta^1=\partial_0\tilde\zeta^1-w^{-1}\zeta^0$.
Then the longitudinal modes $\zeta^0$ and $\tilde\zeta^1$
again have the same action as the ghosts,
leaving us with three massive and five massless
transverse oscillations.

The same conclusion is reached by starting with the Nambu form
of the action and choosing static gauge,
where $r$ and $\phi$ in \asdccd\ are
identified with the world-sheet coordinates.
Let us denote $\xi^s,\xi^4,\xi^q$ the fluctuations of the $x^s$, $w$
and the $S^5$ coordinates respectively. After rescaling
$\bar\zeta^s=w^{-1}\xi^s$, the action is
(here $g_{ij}$ is the induced metric)
\eqn\asddfd{
S=\ha\int d^2\s\sqrt{g}
\left(
g^{ij}\partial_i\bar\zeta^s\partial_j\bar\zeta^s+2\bar\zeta^s\bar\zeta^s
+g^{ij}\partial_i\xi^4\partial_j\xi^4+2\xi^4\xi^4
+g^{ij}\partial_i\xi^q\partial_j\xi^q
\right)\,,
}
and the fields are normalized as
\eqn\norc{
\|\xi\|^2
=\int dr\,d\phi\sqrt{g}\left(
\bar\zeta^s\bar\zeta^s
+{1\over w^2}\xi^4\xi^4
+\xi^q\xi^q
\right).
}

Note that the field $\xi^4$ (and $\zeta^4$ above) is not
normal to the surface, but $\tilde\zeta^4$ is.
As explained in Section 3.3, this choice of gauge has a non trivial
ghost determinant \ghod. In the present
case
$ h^\parallel_{ij}= \diag ( 1/w^2, r^2/w^2)$,
so that
\eqn\finfpc{
\Delta_{\rm gh}
= \det\;\!\!^{ 1/2} ( h^\parallel_{ik} h^{kj} )
=\det\;\!\!^{ 1/2} w^2 \,.
}
The most natural way to regularize this determinant is by
redefining the norm of the $\xi^4$, thus removing the extra
normalization factor in \norc. Then the result for the
partition function in this gauge will be identical to the
conformal gauge expression apart from the contributions of
the ghosts and the longitudinal modes.\foot{If we were to
expand the action without fixing the gauge
$\zeta^0=\zeta^1=0$, we would
find the same action, but with $\xi^4$ replaced by $\tilde\zeta^4$,
with the canonical normalization. The two longitudinal fluctuations
$\zeta^0$ and $\tilde\zeta^1$ drop out of the action. Then one could
choose the gauge $\zeta^0=\tilde\zeta^1=0$ (this is the gauge used in
\refs{\fgt,\kinar} in the context of the
bent string configuration).
$\tilde\zeta^1$ and $\zeta^1$ are related through a rotation by
an angle $\cos\alpha=w$. This rotation introduces a Jacobian which
exactly cancels the ghost determinant in the former gauge.
It is also easy to show directly that the ghost determinant is
trivial in this gauge.}

Before gauge fixing, the fermionic Lagrangian is \frmo
\eqn\asddcf{
L_{2\rm F}
=
-i\left(\sqrt{g}g^{ij}\delta^{IJ}-\epsilon^{ij}s^{IJ}\right)
\bar\theta^I\rho_i D_j\theta^J
.}
To put this action in a 2-d covariant fermionic
form in terms of the zweibein and spin connection of the
induced metric we apply a local
$SO(1,9)$ rotation to transform the projected Dirac matrices
$\r_i$ into constant Dirac matrices contracted with the induced
zweibein as discussed in section 3.4. We get (cf. \sett,\cov)
\eqn\asdcf{\eqalign{
D_i\theta^J
&=\DD_i\theta^J-{i\over2}\epsilon^{JK}\rho_i\theta^K,
\cr
\rho_0
&
=\eta_0^{\,a}\Gamma_a
={r\over w}\Gamma_0
=e_0^{\alpha }S\Gamma_{\alpha } S^{-1},
\cr
\rho_1
&
=\eta_1^{\,a}\Gamma_a
={1\over w}\Gamma_1-{r\over w^2}\Gamma_4
=e_1^{\alpha}S\Gamma_{\alpha} S^{-1},
\cr
\DD_0
&=\left(\partial_0+\ha \Gamma_{01}-{r\over 2w}\Gamma_{04}\right)
=S\, \hat\nabla_0 S^{-1},
\cr
\DD_1
&=\left(\partial_1 -{1\over 2w}\Gamma_{14}\right)
=S\, \hat\nabla_1 S^{-1}.
}}
The rotation matrix is
\eqn\sisrot{
S=\exp\left({\alpha\ov 2}\Gamma_{14}\right),
}
with the same angle $\alpha$ as in \rotate. $\hat\nabla_i$ is the
covariant derivative with spinor world-sheet connection,
\eqn\nabcir{
\hat\nabla_0=\partial_0+{1\over 2w}\Gamma_{01}\,,
\qquad
\hat\nabla_1=\partial_1\,.
}
It is therefore natural to transform
 $\theta^I$ to the new variable
$\Psi^I$
\eqn\asdfc{
\theta^I=S\Psi^I\,.
}
Choosing the gauge $\Psi^1=\Psi^2$, the fermionic Lagrangian
becomes
\eqn\lferco{\eqalign{
L_{2 \rm F}
&=
-2i\sqrt{g}
\left(g^{ij}
\bar\Psi e_i^{\alpha}\Gamma_{\alpha}\partial_j\Psi
+i\bar\Psi\Gamma_{01}\Psi\right)
\cr&
 =
-2i\bar\Psi
\left(-{1\over w^2}\Gamma_0\partial_0
+{r\over w}\Gamma_1\partial_1
+i {r\over w^3}\Gamma_{01}\right)\Psi\,.
}}
Here the 10-d Dirac matrices $\G_\a$ play the role of world sheet
Dirac matrices, as we can choose a representation in terms of the
Pauli matrices
$\Gamma_0=i \sigma_2\times I_8$,\ $\Gamma_1= \sigma_1\times I_8$.
As in the case of the straight string \fuum, this is the action
for a spinor of mass $\pm1$ in $AdS_2$.

Another natural way to fix the $\kappa$ symmetry used
in \refs{\kr,\kt}\foot{
$\Gamma_{0123} = i \g_4 \times I_4 \times I_2$
in the notation of \mt,
where 10-d Dirac matrices are represented as
$ \G^a = \g^a \times I_4 \times \s_1$, $a=0,1,2,3,4$.
}
\eqn\one{
\t^1 = \Gamma_{0123} \t^2,
\qquad
{\rm i.e. }
\qquad
\theta^1 = i \Gamma_4 \theta^2
\,.}
This gauge leads to the {\it same} result for the action as
the $\theta^1=\theta^2$ gauge as we shall explain
below.\foot{Note that in the straight string case \strlin\
this gauge is degenerate.}
Expressing $\t^2$ in terms of $\t^1\equiv \t$ one can check that
\eqn\xxxc{
D_i \theta^1
= {1\over\sqrt w}
\left(\del_i + \ha\Gamma_{i1} \right)
\left(\sqrt {w}\,\theta\right)
\,,}
and , in terms of
\eqn\vvvth{
\vartheta\equiv \sqrt{w}\theta\,,
}
the Lagrangian is
\eqn\lagtbc{
L_{2\rm F}
=-2i\bar\vartheta\left[
-\left({1\over w^3}\Gamma_0-{r\over w^3}\Gamma_4\right)
\left(\del_0+\ha\Gamma_{01}\right)
+{r\over w}\Gamma_1\del_1
\right]\vartheta\,.
}
To simplify this expression, we again use a rotation, this time
in the 0-4 plane. Define
\eqn\psic{
\psi=\exp\left({\beta\ov2}\Gamma_{04}\right)\vartheta\,, }
where
\eqn\coshb{
\cosh\beta={1\over w}\,,
\qquad
\sinh\beta={r\over w}\,,
\qquad
{d\beta\over dr}={1\over w^2}\,.
}
Then
\eqn\lagtpc{
L_{2\rm F}
=-2i\bar\psi\left[
-{1\over w^2}\Gamma_0\left(\del_0+\ha\Gamma_{01}\right)
+{r\over w}\Gamma_1\del_1
-{r\over w^3}\Gamma_{104}
\right]\psi
\,.}
Though this action looks different from \lferco, it also describes
a fermion of mass $\pm$1 (the mass term $\Gamma_{104}=i\Gamma_{23}$
commutes with $\Gamma_0$ and $\Gamma_1$, but is antihermitian,
and its square is 1).

The normalization of $\psi$ is
\eqn\norpc{
\|\psi\|^2
=\int dr d \phi \ \sqrt g\ w^{-1}\ \bar\psi\psi\ ,
}
which is different from normalization of $\Psi$ in \lferco.
Like in the bosonic case, the difference of the normalizations
of the fields in the two $\k$-symmetry gauges may be attributed
to the difference in the corresponding ghost determinants.
Indeed, if the gauge condition is $\t^1 = H \t^2$ where
$H$ is some matrix ($H=1$ and $H= i \G_4$ in the two gauges
discussed above), then it is easy to find the ghost determinant
corresponding to the transformation \kapc. In the cases we are
interested in the case where the $\bar x$ background is constant
on $S^5$ (i.e. when $\td \r_i =\r_i$)
\eqn\kaap{
\delta_\k \t^1 = \r_i^- k^{i } \ , \ \ \ \
\delta_\k \t^2 = \r_i^+ k^{i } \ ,
 }
where $ \r_i^\pm = (g_{ij} \pm e_{ij}) \r_j$,
where $e^{ij} = {\ep^{ij}\ov \sqrt g} $
and $k_i$ are unconstrained vector-spinor parameters,
normalized as
$\|k_i\|^2
=\int d^2 \s \ \sqrt g\ g^{ij} \bar k_i k_j$.
Then the ghost determinant is the inverse square root of the
determinant of the spinor matrix\foot{The ghost determinant
can be obtained from the path integral
$$\Delta_{\rm gh}
\int [dk_i] \exp \bigg[ -\int d^2 \s \sqrt g\ \bar k^i
 (\r_i^- - \r_i^+ H^\dagger) (\r_j^- - H \r_j^+) k_j \bigg] =1\
 . $$}
\eqn\ghas{
 g^{ij} (\r_i^- - \r_i^+ H^\dagger) (\r_j^- - H \r_j^+)
\ . }
Since $\r_{(i} \r_{j)} = g_{ij}$ this matrix is a trivial constant
in the $\t^1=\t^2$ gauge when $H=1$, but, in general, it will
depend on the components of the metric (i.e. on the
coordinates) when $H\not=1$. In the gauge $\t^1 = i \G_4 \t^2 $
the resulting local ghost determinant should ``compensate"
(in an appropriate regularization scheme) for the difference in
normalizations of the spinors $\Psi$ in \lferco\ and $\psi$ in
\lagtpc, explaining the equivalence of the results in the two
$\k$-symmetry gauges.

\subsec{\bf Partition function}

We thus end up with the same partition function \arti\
as in the straight string case, i.e. with that of a
supersymmetric field theory on $AdS_2$. The result
is again a constant whose precise
value depends on regularization and measures for the fields.

This should not come as a surprise, since the circle and the straight
line are related by a special conformal transformation, and
the minimal surfaces also transform into each other. This
is not to say that the partition functions should be identical, there
is a subtle difference. Indeed, already the classical actions
for a circle and a straight line are different. The reason can
be traced to the inclusion of the point at infinity. The same
subtlety should be present at the level of 1-loop partition function.
In the case of the straight line it is
natural to work with the strip model for $AdS_2$,
while for the circle, the Poincar\'e disk is more natural.
In calculating the determinant for the former we should
include also functions that do not behave well at infinity,
while in the circle case those should not be included.
It is therefore probable that the calculation in Appendix
B is more appropriate for this problem rather than the
straight string case.

\newsec{`Parallel Lines'}

Our main interest is in the minimal surface ending at the
boundary of \adss\ which is related to the correlation
function of two anti-parallel Wilson loops. The minimal
surface was constructed in \mal\ and accounts for the
leading large $\l$ behavior ($c_0 \sqrt \l\ov L$)
of the ``quark -- anti-quark'' (W-boson) potential in
$\N=4$ SYM theory. The
first correction $c_1\ov L$ to the potential will be given by the
one-loop partition function of the type we study here.
While some aspects of this computation were addressed before
\refs{\kt,\fgt,\kinar}, our aim below will be to clarify
some previously encountered problems and to set up a
systematic framework which should allow
to compute the finite numerical coefficient $c_1$.

\subsec{\bf The classical solution}
In this section we will write the \adss\ metric in terms of the
coordinate $y=w^{-1}$ (cf. \adsmet)
\eqn\ads{
ds^2 = R^2 (y^2 dx^n dx^n + { dy^2 \ov y^2} + d \Omega_5^2) \,.
}
Here $n=0,1,2,3$ and we will use the index 4 to label
the coordinate $y$. We will often
 set the radius $R$ to be 1 in what follows.\foot{To make the flat
space limit explicit
one should define the coordinate $\vp$ related to $y$ by
$y = R\inv e^{- \vp/R}$. Then
$ds^2 = e^{-2\vp/R} dx^n dx^n + { d\vp^2} + R^2 d \Omega_5$
which becomes flat in the $R\to \infty$ limit.}
If the Wilson lines are extended in the $x^0$ direction and located at
$x^1=\pm {L\ov 2}$, the minimal surface is given by a function $y(x^1)$
(we use world-sheet coordinates $\s^i = x^i = (\tau,\sigma)$, \ $0<\tau < T$).

The bosonic part of string action is then ($y' \equiv \del_1 y $)
\eqn\acc{S= { R^2 \ov 2 \pi \a'} T \int d\s \sqrt { y'^2 + y^4}\ .
}
The stationary point is determined by the second-order equation
$ y y'' = 4 y'^2 + 2 y^4 $
 with the first integral being
 \eqn\fiyt{
y'^2 = {y^8\ov y_0^4} - y^4\ .
}
$y_0$ is an integration constant, the minimal value of $y$.
The special case of $y_0=0$ corresponds to the
``straight string'' configuration discussed in Section 4.
This special solution is a useful reference point:
near the boundaries of the $\s$-interval it gives a good
approximation to the general solution.

The induced metric is
\eqn\stind{
g_{ij}=\pmatrix{y^2 & 0 \cr 0 & y^6},
\qquad
\sqrt g=y^4\,,
\qquad
\sqrt g g^{ij}=\pmatrix{y^2 & 0 \cr 0 & y^{-2}}.
}
The distance between the quark and anti-quark $L$ is
related to $y_0$ by \mal
\eqn\orms{
L= \int^{L/2}_{-L/2} d\s = 2 \int_{y_0}^\infty { dy\ov y^2 \sqrt{
{y^4\ov y^4_0 } -1} }
 = {\k_0\over y_0}
\,, }
\eqn\dedd{
\k_0 \equiv {(2\pi)^{3/2}\ov [{\Gamma({1 \ov 4})}]^2} \ . }
We shall often set $y_0 =1$ as the
dependence on this parameter can be easily restored by
rescalings ($\tau \to y_0\inv \tau$, i.e.
$T \to {T \ov y^2_0}$).
Then
\eqn\yy
{y'^2 = y^8 - y^4,
\qquad
y''= 4 y^7 - 2 y^3
.}

Let us first review how the classical contribution is computed.
The action \acc\ takes the following value on the solution
\eqn\icc{S=
 { R^2 T \ov 2 \pi \a' y^2_0 }
\int^{L/2}_{-L/2} d\s \ y^4\ . }
Since
$ (y^{-3}y')' = {y^4 + y_0^4\ov y^2_0 } $
is a total derivative (which goes to the boundary
where $y=\infty$ and gives only a trivial divergence),
we can replace $y^4$ by $-y^4_0$,
assuming that the infinite boundary contribution should be dropped.
This prescription is the same as normalizing
the partition function to the straight line case
(and is essentially equivalent to
the one in \dgo: the Legendre transform subtracts
the same boundary term or total derivative\foot{
This is an example of an amusing relation. The Legendre
transform can be written as an integral over a (rather
complicated) total derivative. Instead, one can note that for
smooth loops the Legendre transform which is equal to the
divergence in the area is also equal (asymptotically) to the
geodesic curvature $K$. Then we can use the Gauss-Bonnet
theorem to write the action as
$S={R^2\over2\pi\a'}\int d^2\s\sqrt{g}\,(1+{1\over 2}\R)
-{R^2\over\a'}\chi$, where $\chi$ is the Euler number.
Since $\R$ approaches $-2$, the integral is manifestly
convergent.
For the present geometry $\chi=0$.
}).
Then
\eqn\icyc{S=
 - { R^2 T \ov 2 \pi \a'} y^2_0
\int^{L/2}_{-L/2} d\s =
 - { R^2 T L \ov 2 \pi \a'} y^2_0
= -
{ \l^{1/2} (2 \pi)^{2}\ov [\Gamma({1 \ov 4})]^{4}} { T \ov L}\
 , }
where $\l = 4\pi g_s N = {R^4 \ov \a'^2}$.
This result is the same as in \mal, here
found in `one shot' (without doing any further integrals).

In the flat space limit ($R\to \infty$) one finds that the
quantum correction \aca\ to the potential vanishes because of
the cancellation of the bosonic and fermionic contributions due
to effective 2-d supersymmetry present after gauge fixing
(see Section 2.3). As was pointed out in \kt, the 1-loop
$c_1 \ov L$ correction to the effective potential
may not, however, vanish in the present curved space case
as there is no reason to expect that the action expanded
near the solution $y=y(\s)\not=\const$ should have an
effective world-sheet supersymmetry.

\subsec{\bf Quadratic fluctuations: bosons}

In conformal gauge the bosonic action is \ctoq, where $X_{pq}=0$
and\foot{In this form of the metric \ads\ the vielbein components are
$E^m_n = y\d^{m}_n$, $E^4_4 = y^{-1}$.
If we restore the $R$ dependence,
$X_{ab} \to {1 \ov R^2}X_{ab}$, then the ``mass term''
vanishes in the flat space limit as it should.}
\eqn\rewq{\eqalign{
&X^{ab}=2\delta^{ab} - g^{ij}\eta_i^{\,a}\eta_j^{\,b}\,,
\qquad
\eta_0^{\,a} = (y, 0, 0,0,0)\,,
\qquad
\eta_1^{\,a} = (0, y, 0,0,y\inv y')\,,
\cr
&
D_i \z^a = \del_i \z^a + w^{ab}_i\z^b,
\qquad
w^{ab}_i= \del_i x^\m \Omega^{ab}_\m,
\qquad
w^{a4}_i = y \d^{a}_i = - w^{4a}_i
.}}
The ghost action is \ghostqq\ or \ghostc\ where
the covariant derivative $\nabla_i$ includes the world-sheet
spin connection $\omega_0^{01}=y^{-3}y'$. The action contains
the curvature $\R$ of the induced metric $g_{ij}$
\eqn\cur{
\sqrt g \R = \left(1\over y^2\right)'',
\qquad
\R = -2 \left(1 + {1 \ov y^4}\right)
. }
Unlike the circle case, here there is no obvious rotation
of the fields $\z^a$ such
that the contribution of the longitudinal modes becomes the same as
that of the
ghosts.\foot{In fact, the eigenvalues of the mass matrix $X^{ab}$
are $(1,1,2,2,2)$, not $(-\ha \R,-\ha \R,4+\R,2,2)$. In addition
there are extra connection terms that remain after the rotation.}

Choosing the static gauge in the Nambu action
 we denote by $\xi^s$ ($s=2,3$) the fluctuations of
the two ``longitudinal'' 3-brane directions, by $\xi^4$ the
fluctuation along the radial $y$-direction, and by $ \xi^q$
($q=5,...,9$) the fluctuations in the 5-sphere directions. Their
natural norms are
\eqn\nor{
\|\xi\|^2 = \int d^2 \s \sqrt g ( y^2 \xi^s \xi^s + y^{-2} \xi^4 \xi^4 +
\xi^q \xi^q )\ . }
Introducing the rescaled fields\foot{It should be noted that
redefinitions we make are accompanied by Jacobians and thus do
not introduce new quadratic or linear divergences.}
\eqn\resc{
\zeta^s = y \xi^s,
\qquad\ \ \
\zz^4 = y^{-3} \xi^4,
}
one finds (after integration by parts and use of the properties of
the classical background \yy) the following expression for the
quadratic fluctuation part of the gauge-fixed action\foot{As before,
we absorb $1 \over 2\pi \a'$ factor in the action into a redefinition
of the fluctuation fields.}
\eqn\qua{
S_{2 \rm B} = { 1 \ov 2} \int d^2 \s \sqrt g \bigg[
g^{ij} \del_i \zeta^s \del_j \zeta^s + 2 \zeta^s\zeta^s
+g^{ij} \del_i \zz^4 \del_j \zz^4 + (\R + 4) \zz^4\zz^4
+g^{ij} \del_i \xi^q \del_j \xi^q \bigg]
\ . }
As follows from \nor, the fields in \qua\ are normalized as follows
\eqn\norb{
\|\xi\|^2 = \int d^2 \s \sqrt g ( \zeta^s \zeta^s
 + y^{4} \zz^4 \zz^4 + \xi^q \xi^q )\ . }
Thus, while the action $S_{2\rm B}$ seems to have a `covariant' 2-d form
with respect to the induced geometry (two massive scalars, one
scalar with a potential, and 5 massless scalars in external 2-d
metric), this is not true for the measure because of the
$y^4$ factor in the $\zz^4\zz^4$ part. This is remedied by
the inclusion of the ghost determinant \ghod, where
in our case $ h^\parallel_{ij}=\diag ( y^2, y^2)$, so that
\eqn\paragh{
\Delta_{\rm gh} = \det\;\!\!^{ 1/2} ( 1/y^4 )\,.
}
The most natural regularization of this determinant is achieved by
rescaling the field $\zz^4$, which cancels the renormalization
factor in \norb, much as in the case of the circle in
Section 5.1 (see \finfpc).

The same expression for the action \qua\ was found in \fgt. There
instead of $\zz^4$ the authors used the fluctuating field normal
to the surface
$
 \eta^4 \equiv - \sin \a\ \z^{ 1} + \cos \a\ \z^{ 4}
= - y^{-3} y' \xi^{ 1} + y^{-3} \xi^{ 4}\,,
$
where $\a$ is the angle defined by
$\cos \a = y^{-2}, \ \a'=2y$,
and $
\z^{ a} = E^a_{\m} \xi^\m
$
are the target-space vielbein components of $\xi^\m$. The
normalization of $\eta^4$ is canonical, and it is easy to
see that the ghost determinant is trivial in that gauge.

\subsec{\bf Quadratic fluctuations: fermions}

In the present case of the classical solution
\fiyt, which is constant in $S^5$ directions, one finds
that the leading quadratic part of the fermionic action
given by \frmo\ depends on
\eqn\expi{
\r_0 = y \Gamma_0\,,
\qquad
\r_1 = y \Gamma_1 + y\inv y' \Gamma_4\,,
\qquad
{\cal D}_i = \del_i + \ha y \Gamma_i \Gamma_4\,,
}
where we used the fact that for $AdS_5$ space the non-vanishing
components of the connection are
$\Omega^{n4}_\m = E^n_\m, \ n =0,1,2,3$.
Then \frmo\ becomes
\eqn\frm{
L_{2\rm F} = i\left( \sqrt{g} g^{ij}\d^{IJ} - \ep^{ij} s^{IJ}\right)
\left( \bar \t^I \r_i {\cal D}_j \t^J
 - \ha i \ep^{JK} \bar \t^I \r_i \r_j \t^K\right)
\,.}
Here $g_{ij}$ is the Minkowski version of the induced metric \stind,
i.e. the corresponding 2-d vielbein components $e^\a_i$ are
\eqn\vier{
e^{ 0}_0 = y\,,
\qquad
e^{1}_1 = y^3\,,
\qquad
g_{ij}=\diag\left(-y^2,y^6\right)\,.
}

The crucial observation, that allows us to put the action
\frm\ into a simple 2-d
covariant form, is that the combination of $\G_1$ and $\G_4$ which
appears in $\r_1$ can be interpreted as a (local, $\s$-dependent)
rotation of $\G_1$ in the 1-4 plane
\eqn\ccc{
S \G_1 S\inv = \cos \a\ \G_1 + \sin \a\ \G_4
= y^{-2} \G_1 + y^{-4} y' \G_4 = y^{-3} \r_1\,,
}
where
\eqn\fff{
S= \exp \left( - {\alpha\over 2}\G_1 \G_4\right)\,,
\qquad
\cos \a = y^{-2}\,,\
\ \
\sin \a = y^{-4} y'\,,
\ \ \
\a'\equiv {d\a\ov d \s} = 2 y\,.
}
Making the field redefinition
\eqn\uuu{
\t^I \to \Psi^I \equiv S\inv \theta^I\,,
}
we then find that \frm\ takes the following simple `2-d covariant'
 form
\eqn\frmk{
L_{2\rm F} = i\left( \sqrt{g} g^{ij}\d^{IJ} - \ep^{ij} s^{IJ}\right)
 \left( \bar \Psi^I \tau_i {\hat \nabla}_j \Psi^J
 - \ha i \ep^{JK} \bar\Psi^I \tau_i \tau_j \Psi^K\right)
\,,}
where $\tau_i$ play the role of
the curved space 2-d Dirac matrices\foot{To prove eq.\frmk\ one
should note that
$\tau_i = S\inv \r_i S$ and that
$\td {\cal D}_i \equiv S\inv \DD_i S$ is found to be (see \fff)
$\td \DD_0 = \del_0 + \ha y\G_0 ( y^{-4} y' \G_1 + y^{-2} \G_4 )
= \hat \nabla_0 + B_0$,
$\td \DD_1 = \del_1 - y \G_1 \G_4 + \ha y \G_1 \G_4 = \nabla_1 + B_1$,
where
$B_0 = \ha y^{-2} \tau_0 \G_4$, $B_1 = -\ha y^{-2} \tau_1 \G_4$.
Finally, one observes that the connection $B_i$ drops out from the
action since $\tau^i B_i=0$,\ $\ep^{ij} \tau_i B_j =0$.}
\eqn\ttrr{
\tau_i = e_i^{\a} \G_{\a}\,,
\qquad
\tau_0 = S\inv \rho_0 S = y \G_0\,,
\qquad
\tau_1 = S\inv \rho_1 S = y^3 \G_1\,,
}
and $\hat \nabla_i$ is the 2-d curved space spinor covariant
derivative corresponding to \vier\
\eqn\gty{ \hat \nabla_k =\del_k + \four \omega^{\a\b }_k
\Gamma_{\a \b}\,,
\qquad
\hat \nabla_0 = \del_0 + \ha y^{-3} y' \G_0 \G_1\,,
\qquad
\hat \nabla_1 = \del_1\,.
}
The Lagrangian \frmk\ is then
\eqn\fum{
L_{2\rm F} = i \left( \sqrt{g} g^{ij} - \ep^{ij} \right)
 \bar \Psi^1 \tau_i {\hat \nabla}_j \Psi^1
+ i \left( \sqrt{g} g^{ij} + \ep^{ij} \right)
 \bar \Psi^2 \tau_i {\hat \nabla}_j \Psi^2
- \ep^{ij} \bar\Psi^1 \tau_i \tau_j \Psi^2
\,.}
It is easy to see that the covariant derivative and
`mass' terms here are separately invariant
under the leading-order $\k$-symmetry transformations
 (see \kapc, \kaap)
$\delta_\k \t^I = \r_i \k^{i I}$
or their `rotated' form
\eqn\uiy{ \delta_\k \Psi^I = \tau_i \k'^{i I}\,,
\qquad
\k'^{i I} = S\inv k^{iI}\,.
}
Fixing the
$\kappa$-symmetry gauge by the condition
\eqn\gue{ \t^1=\t^2\,,
\qquad
{\rm i.e. }
\qquad
 \Psi^1 =\Psi^2\equiv \Psi \,,
}
 we get
\eqn\fuumqaq{
L_{2\rm F} = 2i \sqrt{g} \left( \bar \Psi \tau^i {\hat \nabla}_i \Psi
 + i \bar \Psi \tau_3 \Psi \right)
\,,}
\eqn\you{
 \tau_3 \equiv {\ep^{ij} \ov 2\sqrt{g}} \tau_i \tau_j =
\G_0 \G_1\,,
\qquad
(\tau_3)^2 =1\,.
}
Note that choosing this gauge before the rotation, the
action \frm\ may be written as
\eqn\wrir{
L_{2\rm F} = 2 i \sqrt{g} \bar \t \r^i \Dd_i \t
= i \sqrt{g} ( \bar \t \r^i \del_i \t
- \del_i \bar \t \r^i \t + i e_{ij} \bar \t \r^i\r^j \t )
\ , }
where $\Dd_j = {\cal D}_j + \ha i e_{jk} \r^k,$\
 $\r^i = g^{ij} \r_j$,\ \
$e_{jk} \equiv { 1 \ov \sqrt g} g_{jj'} g_{kk'} \ep^{j'k'}$.
To see that the rotation \ccc,\fff\ is indeed a special case of
the general rotation \sett\ discussed in Section 3.4 we note
that here the tangent $t^\m_\a$ and normal $n^\m$ vector components
are (here $\m,\n=0,1,4$ label the target space $AdS_3$ coordinates
inside $AdS_5$, i.e. $G_{\m\n} = (y^2, y^2, y^{-2}$)):
\eqn\hjjhh{
 t^\m_{\hat 0}= (1,0,0)\ , \ \ \
t^\m_{\hat 1}= (0, y^{-3}, y^{-3} y')\ , \ \ \
n^\m = (0, - y^{-5}y', y^{-1})\ .
}
Then $\r_\a = (\r_{\hat 0}, \r_{\hat 1})
$ and $\r_s\equiv \r_\perp$ are (cf. \ccc)
\eqn\roten{
\r_{\hat 0 } = \G_0 \ , \ \ \
\r_{\hat 1 } = y^{-3} \r_1 = y^{-2} \G_1 + y^{-4} y' \G_4 \ ,
\ \ \
\r_{\perp } = y^{-3} \r_1 = - y^{-4} y' \G_4
 + y^{-2} \G_4 \ ,
}
so that
\eqn\hfhfgh{
- S\inv dS = { 1 \ov 4} ( \r_\a d \r^\a + \r_\perp d \r_\perp)
= \ha \G_1 \G_4 d \a = y \G_1 \G_4 d\sigma \ ,
}
in agreement with \fff.
This is $U(1)$ rotation that does
not lead to a non-trivial Jacobian
in the present case of non-chiral 2-d fermions.

Another possible gauge choice is the analog of the covariantized
light cone gauge of \wig\ (see \trw,\expi):
\eqn\covl{
(\r_0+\r_1) \Psi^1 =0\,,
\qquad
(\r_0-\r_1)\Psi^2 =0 \,,}
or, explicitly, after the rotation \ttrr,\foot{Note that this gauge choice
is different from the one in \fgt\ where instead of $\tau_\pm$
the combinations $\G_0\pm \G_1'$ ($\G_1'= S \G_1 S\inv$) were used
(the rotation was not explicitly done in \fgt).}
\eqn\covl{
\tau_+ \Psi^1= (y\G_0+ y^3\G_1) \Psi^1 =0 \,,
\qquad
\tau_- \Psi^2= (y\G_0- y^3\G_1) \Psi^2 =0 \,.}
The resulting action is essentially
the same as \fuumqaq\ with left and right parts
of $\Psi$ explicitly separated.

Choosing a representation for $\G_a$ such that
$\G_{0,1}$ are 2-d Dirac matrices times a unit $8\times 8$
matrix, i.e.\foot{Recall that
the original $32\times 32$ Dirac matrices are such that
$\G_a= \g_a \times I_4 \times \s_1$, or simply
$\G_a= \g_a \times I_4$ on a 16-subspace
of left MW spinors, with $\g_{(a}\g_{b)} = \diag(-1,+1,+1,+1,+1)$.
We are not distinguishing between $\G_a$ and $\g_a $,
i.e. we treat $\Psi$ as a 4-component spinor suppressing
its extra 4 spectator indices. }
\eqn\gde{ \G_0 = i \s_2 \times I_8\,,
\qquad
\G_1 = \s_1 \times I_8\,,
\qquad
\tau_3= \G_0 \G_1 = \s_3 \times I_8\,,
}
where $\s_{1,2,3}$ are the Pauli matrices, we end up with 8
species of 2-d Majorana fermions living on a curved 2-d surface
with a $\s_3$ mass term. {Assuming that fermions are normalized
with $\sqrt g$}, the square of the resulting fermionic operator is
then ($\tau_3^\dagger =\tau_3$)\foot{Note that Dirac operator is
self-adjoint with the measure $\sqrt g$.}
\eqn\fermi{\Delta'_F
= (D_F)^2 =
 ( i \tau^i {\hat \nabla}_i - \tau_3 )
 ( i \tau^j {\hat \nabla}_j - \tau_3 )
= - \hat \nabla^2 + \four \R + 1 \,,
}
where $ \hat \nabla^2 = { 1 \ov \sqrt g} \hat \nabla^i
(\sqrt g \hat \nabla_i).$\foot{Since
$\tau_i = e^{\a}_i \G_{\a}$, where $\G_{\a}$ are {\it constant},
and since $\hat \nabla $ is the covariant 2-d spinor derivative,
the squaring relation is exactly the same as for the standard
2-d fermions in curved space.}
Explicitly (recall that here we use the Minkowski signature)
\eqn\dirr{\eqalign{
\Delta'_F
&=\, - \left[y\inv \left(\G^0 \del_0 + \ha y^{-3} y' \G_1\right)
+ y^{-3} \G^1 \del_1 \right]^2 + 1
\cr&
=\,
y^{-2} \left(\del_0 + \ha y^{-3} y' \G_0 \G_1\right)^2
- y^{-4} \del_1 \left(y^{-2} \del_1\right)
+ \ha \left( 1 - y^{-4}\right) \,.
}}
A similarly looking result for the fermionic operator was found
in \fgt\ where a different $\k$-symmetry gauge was
used.\foot{The kinetic part of our operator is actually different
from the expression in \fgt\ which contained an additional connection
term in the covariant derivative, and our derivation of the
action is much more straightforward. After submission of this paper 
S.~F\"orste pointed out to us that the two expressions might be 
related to each other by a rotation, i.e. are equivalent.}

One can also consider the quadratic fermionic action \frm\ in the
``3-brane'' gauge $\t^1= i \G_4 \t^2$ \one.
It was found in \kt\ that the sum of the quadratic fermionic
terms in the action of \mt\ takes the simple form
\eqn\kat{\eqalign{
S_{2\rm F}
=&\,
2i \int d^2\sigma
\left(\sqrt g g^{ij} y^2 \bar \vt \Gamma_i \partial_j\vt
- \ep^{ij} \partial_i y \bar \vt \G^4
\partial_j\vt\right)
\cr
=&\,
2i \int d^2\sigma\, (y^{-1/2} \bar \t) \left[
\Gamma^1 \partial_1 +
( y^4 \Gamma^0 + y' \G^4) \partial_0 \right] (y^{-1/2} \t) \,,
}}
where
$\t\equiv\t^1$ is the original GS target space spinor
variable related to the rescaled field $\vt$ in \kt\
by $\t= y^{1/2} \vt$.
Here $\G^a$ are constant Dirac matrices
($\G^0= -\G_0$, $\G^m=\G_m$, $m=1,4$).
As was noted in \kt, the resulting fermionic operator is
non-degenerate.

At first sight, this operator is very different from the one in
\fuumqaq; in particular, it has no mass term. But the two are,
in fact, closely related! To demonstrate this let us note that
the combination of the $\G$-matrices multiplying $\del_0$ is
actually a local Lorentz rotation of $\G^0$ in the 0-4 plane
with parameter $\b$
\eqn\ddd{
\S= \exp \left( - {\beta\over 2} \G^0 \G^4\right)\,,
\qquad
\cosh \b = y^{2},
\qquad
\sinh \b = y^{-2} y'\,,
\qquad
\b'= 2 y^3\ ,
}
i.e.
\eqn\eqnie{
\S \G^0 \S\inv = \cosh \b\ \G^0 + \sinh \b\ \G^4
 = y^{2} \G^0 + y^{-2} y' \G^4\ .
}
Introducing
\eqn\ccch{
\chi = \S\inv \t \,,
}
we get (note that $\bar \t = \bar \chi \S\inv$ for any $SO(9,1)$
rotation)
\eqn\get{
L_{2\rm F} = 2i \bar \chi \left[
 y\G^0 \del_0 + y\inv \G^1 \del_1 - \ha y^{-2} y' \G_1
+ y^2 \G^0\G^1\G^4
\right] \chi
\,. }
To put this action into the `curved space 2-d spinor'
form we need to make a redefinition $\chi\to \psi$
\eqn\yti{
\chi = y \psi \,,
}
\eqn\sssd{
L_{2\rm F} = 2i \bar \psi
\left[ y^3 \G^0 \left(\del_0 + \ha y^{-3} y' \G_0\G_1\right)
+ y \G^1 \del_1 + y^4 \G^0\G^1\G^4\right] \psi
\,.}
Since $i\G_4 = \G_0 \G_1\G_2\G_3$, i.e.
\eqn\nonnn{
\G^0\G^1\G^4 = i \G^2 \G^3,
}
we finally get, using \vier,\gty, the expression that is essentially
{\it equivalent} to \fuumqaq
\eqn\fuom{
L_{2\rm F} = 2i \sqrt{g} \left( \bar \psi \tau^i \hat \nabla_i \psi
 + i \bar \psi \tau'_3 \psi \right)\,,
\qquad
\tau_3' \equiv \G_2 \G_3\,,
\qquad
\left(\tau_3'\right)^2 =-1 \,.
}
The square of the fermionic operator in \fuom\ is indeed the same
as in \fermi:
\eqn\ermi{
\Delta'_F = (D_F)^\dagger (D_F) =
\left( i \tau^i {\hat \nabla}_i + \tau'_3 \right)
\left( i \tau^j {\hat \nabla}_j - \tau'_3 \right)
= - \hat \nabla^2 + \four \R + 1 \,,
}
where we used that, in contrast to $\tau_3$ in \fuumqaq,
$\tau_3'$ is {\it anti}hermitean and {\it commutes} with
$\tau_i$.

Since $\t$ and thus its image under the
rotation $\chi$ are assumed to have canonical
normalization, $\|\chi\|^2 =\int d^2 \s \sqrt g \bar \chi \chi$,
we conclude that $\psi$ in \yti\ should
be normalized with the extra factor of $y^2$.
As in the case of the circle in Section 5.1, this extra
factor is offset by the non-trivial
$\k$-symmetry ghost determinant \ghas\
corresponding to the gauge $\t^1 = i \G_4 \t^2$,
so the fermionic contributions
in the two $\k$-symmetry gauges are again equivalent.

\subsec{\bf Partition function}

Let us first combine the bosonic contributions. In the
conformal gauge
\eqn\partic{
Z_{\rm bose, \ conf.g.}
= { \det^{1/2}\left( - \nabla^2_{ij} - \ha \R g_{ij} \right)
\over
\det^{1/2} \left( - D^2_{ab} + X_{ab} \right)
\det^{5/2} \left( - \nabla^2 \right) }\,,
}
while in the static gauge
\eqn\partit{
Z_{\rm bose, \ stat. g.}
= { 1 \over \det^{2/2} \left( - \nabla^2 + 2 \right)
\det^{1/2} \left( - \nabla^2 + \R + 4 \right)
\det^{5/2} \left( - \nabla^2 \right)}
\,.}
Up to global factors in the gauge group,
the two expressions must be equivalent; it is easy to see,
for example, that the corresponding logarithmic
divergence coefficients are indeed the same
\eqn\vanik{
(b_2)_{\rm bose, \ conf.g.} =
 (5+5 -2) \times { 1 \ov 6} \R - 8 - \R\ ,
}
\eqn\anik{
(b_2)_{\rm bose, \ stat. g.}
 = (2+1+5) \times { 1 \ov 6} \R - 2 \times 2 - 4 - \R\ ,
}
where we used \rewq. The contributions of the massless determinants
and the ghost determinant can be found, as usual, by integrating the
conformal anomaly (see Appendix A), but to find the massive
determinants one needs to solve the corresponding spectral problems.

Including fermions, the expression for the 1-loop partition
function of a string in \adss\ background with world surface
ending on two parallel lines is thus
\eqn\parti{
Z_{AdS_5 \times S^5}^\parallel
= { \det^{8/2} ( - \hat \nabla^2 + \four \R + 1)
\over \det^{2/2} \left( - \nabla^2 + 2 \right)
\det^{1/2} \left( - \nabla^2 + \R + 4 \right)
\det^{5/2} \left( - \nabla^2 \right)}
\,,}
which is essentially the same as in \fgt.
Here we took the fermionic contribution in the $\t^1 =\t^2$
gauge,
and the bosonic contribution in the static gauge.

The geometry under consideration is asymptotic to $AdS_2$.
For example, if we change the coordinate $\s$ to $y$
the induced metric takes the form\foot{It is sometimes useful
to use the coordinate $w=1/y$, in terms of which the metric is
$ds^2 = { 1 \ov w^2} [ d\tau^2 + { w_0^4 \ov w_0^4 -w^4} dw^2 ]
$,
where $0< w< w_0= 1/y_0$ and $w=0$\ corresponds to the boundary.}
\eqn\hhu{
ds^2 = y^2 d\tau^2 + { y^2 \ov y^4 -y_0^4} dy^2 \ ,
\ \ \ \ \ \ \ y_0 \leq y< \infty \ . }
The $y_0=0$ limit of \hhu\ corresponds to the straight string
configuration where the metric becomes that of Euclidean
$AdS_2$ space (with $0 \leq y< \infty$).

As usual in the case of negative curvature non-compact spaces
similar to $AdS$, the corresponding actions will in general
be divergent and one will need to add boundary counterterms.
The details of how the divergent boundary behavior is properly
accounted for is actually rather irrelevant for the purpose of
extracting the non-trivial finite $T\ov L$ part of
$\ln Z_{AdS_5 \times S^5}^\parallel$ we are interested in.

To avoid altogether questions about boundary terms (and details
of topological infinity cancellation) we may normalize our
partition function for each field by the partition function of
an equivalent field in the straight
string configuration, i.e. divide the partition function \parti\
for the noncompact hyperbolic negative curvature space \hhu\ by
the partition function (twice, to account for the two asymptotic
regions), \arti\ for the $AdS_2$ case
\eqn\divi{
\bar Z_{AdS_5 \times S^5}^\parallel (y_0) =
{ Z_{AdS_5 \times S^5}^\parallel(y_0)
\ov Z_{AdS_5 \times S^5}^\parallel(0)} \ .
}
Since the topology and the near-boundary (large $y$) behavior of
the two metrics is the same, this eliminates the problem of carefully
tracking down all boundary terms in the expressions for the
determinants and allows us to ignore the boundary contributions
as well as the total derivative bulk terms (such as the
logarithmically divergent terms proportional to
$\int d^2 \s \sqrt g \R$).\foot{It is easy to see that the
divergent integral $\int d^2 \s \sqrt g \R$ gets contribution only
from the boundary behavior of the metric.}

The ratio of the determinants for the metric \hhu\ and
for its $y_0=0$ limit will be finite and well-defined.
This is actually the standard recipe of defining the
determinants of Laplace operators on (e.g. 2-dimensional)
non-compact spaces by using fiducial metrics of constant
negative curvature which have the same topology and asymptotic \
behavior. As a result, one will need to compute only
the well-defined heat kernels like
$\Tr [e^{- t \Delta(y_0)} - e^{- t \Delta(0)}]$.

>From a practical point of view, the subtraction of the
$AdS_2$ contributions allows us, as in the evaluation
of the classical action \icc,\icyc, to freely integrate
by parts and to drop all divergent boundary contributions.
Some examples illustrating this procedure are given in
Appendix A.3.

\subsec{\bf Crude approximation for the 1-loop potential}

One may make the following very simple (but probably too crude)
estimate of the value of the resulting partition function and
thus of the coefficient in the 1-loop correction $ c_1/L$ to the
potential. The classical solution $y$ as a function of $\s$
is approximately equal to $y_0$ and changes slowly near
$\s\approx 0$ and then blows up to infinity at the boundaries
of the $\s$-interval $( - L/2, L/2)$. It seems reasonable to assume
that the near-boundary behavior of $y(\s)$ should not be very
important for the value of the normalized partition function \divi.
One may then approximate $y(\s)$ to be made of three straight
pieces. A part where $y\approx y_0, \ y'\approx 0,\ y''\approx 0$,
and two parts connecting it to the boundary. Note that this is
{\it not} the same as taking the flat space limit, since now in
\parti\ we have determinants of operators with non-zero mass.

Below we will estimate the contribution of the part at
$y\approx y_0$. We did not evaluate the contribution from the two
pieces connecting it to the boundary.

Since $y$ is assumed to change very slowly, we may set $\R\approx 0$.
Then we are left with the following combination of determinants
in flat metric\foot{Contributions of overall constant factors like
$y^{-2}_0$ in the operators cancel out due to supersymmetric balance
of the numbers of fields.}
\eqn\rati{\eqalign{
W =&\, \ha \bigg[\ 2\ \ln \det (-\del^2 + 2 y^2_0 )
+ \ln \det (-\del^2 + 4 y^2_0 )
+\ 5\ \ln \det (-\del^2 )
\cr&\ \ \
- \ 8\ \ln \det (-\del^2 + y^2_0 )
\bigg]\ .
}}
This effective action is UV finite because of the obvious mass sum
rule. The non-zero finite result for $W$ can probably be interpreted
as the vacuum energy of some spontaneously broken supersymmetric 2-d
field theory corresponding to \rati.

Assuming Dirichlet boundary conditions in both
$\tau$ and $\s$ directions and taking $T\to \infty$
(so that we can integrate over the continuous
eigenvalue in $\tau$-direction) we get\foot{In general,
for a massive determinant we get
$ \prod_{n,k=1}^\infty \big[ ({ \pi n \ov T})^2
+
 ({ \pi k \ov L})^2 + m^2 \big]$.
For a massless determinant
$ \prod_{n,m=1}^\infty \big[ ({ \pi n \ov T})^2
 + ({ \pi m \ov L})^2 \big]
= (2L)^{-1/2} \eta( i{ T\ov L}), $
where $\eta(x)
= e^{ i{\pi \ov 12} x} \prod_{n=1}^\infty
( 1 - e^{ i\pi n x})$.}
\eqn\lop{\eqalign{
& \ha \ln \det (-\del^2 + m^2) - \ha \ln \det (-\del^2)
\cr&\ \ \ \ \ \ \ \
= \ \ha T \sum_{n=1}^\infty
 \int^\infty_{-\infty} { dk \ov 2\pi}\,
 \ln\left(1 + { \om^2_n \ov k^2}\right)
= \ha T \sum_{n=0}^\infty { \om_n} \ , \ \ \ \ \
\om^2_n = ({ \pi n \ov L})^2 + m^2\ .
}}
Thus
\eqn\soo{
W= z_1 { \pi\ T\ov 2\ L}\,,
\ \ \ \ \
z_1= \sum_{n=1}^\infty
\bigg[ \sqrt{n^2 + 4 a^2} + 2 \sqrt{n^2 + 2 a^2}
+ 5 \sqrt{n^2} - 8 \sqrt{n^2 + a^2} \bigg] \ , }
\eqn\noo{
a = {L y_0 \ov \pi}
= { 2 \sqrt{2 \pi}
 \ov [\Gamma({1 \ov 4})]^{2}} \approx 0.38138 \ ,
}
where we have used \orms. To compare, in the ``massless" case
one finds \refs{\LU} (see \aca)
\ $z_1 =\sum_{n=1}^\infty n = \zeta(-1) = - { 1 \ov 12} \approx - 0.08333$.
The infinite sum \soo\ is {\it convergent}
and its the numerical evaluation gives
\eqn\numer{
z_1 \approx - 1.24966\ .
}
Thus the coefficient of the $1/L$ potential is negative, i.e.
has the same sign as a boson in flat space.

To this one has to add the contribution of the two
``flat'' lines connecting it to the boundary. This will give a result that is
not identical to that of $AdS_2$, because these lines extend only up to
$y_0$.

To go beyond the above crude approximation and to compute
$\ln \bar Z_{AdS_5 \times S^5}^\parallel = - c_1 { T \ov L} $
in \divi\ exactly one may use the following strategy:
(i) first, one may compute the contributions of the massless
determinants and the Jacobian using the results of Appendices
A and C;
(ii) then, since the induced 2-d metric is conformally flat,
one may rescale it to the flat space one, isolating the
conformal anomaly parts of the determinants;
(iii) finally, one may compute the spectra of the resulting
operators in flat metric with $y$-dependence being only
in the mass terms. For example, the bosonic operators in
\parti\ then become
\eqn\toops{
-\del^2_0 - \del'^2_1
 + 2 y^2 \ , \ \ \ \
 -\del^2_0 - \del'^2_1 + 2 y^2 - 2 y^{-2} \ , \ \ \
 -\del^2_0 - \del'^2_1
\ , }
where we have made the coordinate change $\s\to \s'$
such that the 2-d metric becomes conformally flat,
\eqn\redd{
 ds^2 = y^2 (d\tau^2 + d\s'^2) \ , \ \ \ \ \ \ \
 d\s' = y^2 d\s \ . }
The computation of the spectra of these operators defined in
the 2-d strip $(T,L')$, where
$L' = { [ \G({1 \ov 4})]^4 \ov 2 (2 \pi)^2 } L$ is the range of
$\s'$ (see Appendix A.3), is left for the future.

\newsec{Conclusions}
We have presented a systematic treatment of the Green Schwarz string
in curved \adss\ space. We found the quadratic fluctuation operators
in conformal and static gauges (for Polyakov and Nambu-Goto actions).
A careful treatment was presented of the measure factors and ghost
determinants.

We also considered two different ways of gauge fixing the
$\k$-symmetry, and explained how one can relate the GS fermion
kinetic term to the standard 2-d Dirac fermion action on a
curved 2-d background by making a local target-space Lorentz
rotation.

We discussed the resolution of the problem of the logarithmic
$O(\R)$ divergences encountered (in the Nambu framework)
in \fgt. First, as in the case of the flat target space,
the divergence proportional to $\int \R$ should be accompanied
by a boundary term, promoting it to the Euler number, and thus
is topological. The cancellation of topological divergences in a
critical string theory is implied by careful definition of the
path integral measure. This is clearer in the Polyakov approach,
but should also be true in the Nambu formulation.
In any case, this issue does not arise in the case of
the induced 2-d geometries that we discussed (except for the
circle), since there the Euler number vanishes.

We have emphasized that the natural way to define the partition
function in the case where the induced geometry is asymptotic to
$AdS_2$ (the case relevant for computing the correction
to quark -- anti-quark potential) is to normalize with respect to
the partition function for $AdS_2$ space. Then the issues of
boundary counterterms and divergences simply do not arise.

We have studied the cases of minimal surfaces ending along a
straight line, a circle and two lines. In the first two cases
we ended up with a supersymmetric field theory on $AdS_2$.

The straight line is a BPS object, and therefore one expects 
$Z=1$, as we were able to verify. We presented two other ways of 
calculating the partition function on $AdS_2$, which give 
different results. The discrepancy is attributed to
different regularizations and to assumptions about the
asymptotics of the eigenfunctions. Those calculations might 
be more appropriate for the circular loop geometry, where 
supersymmetry is broken.

In the case of the two parallel lines we have found the general
expression for the partition function and showed how to
express it in terms of the determinants of 2-d Laplace
operators on a flat strip with potentials depending only on
one of the two coordinates. We have not, however, addressed
the issue of finding exact analytical or numerical methods of
computing the corresponding determinants and thus the value of the
numerical constant in the subleading correction to the
$1/L$ potential.

\bigskip
\bigskip
\centerline{\bf Acknowledgements}
N.D. would like to thank G.~Horowitz, N.~Itzhaki and J.~Polchinski,
and A.A.T. would like to thank R.~Metsaev, P.~Olesen and A.~Polyakov
for useful discussions of related questions. The work of N.D. and
D.J.G. is supported by the NSF under grant No. PHY94-07194.
The work of A.T. is supported by the DOE grant DOE/ER/01545-780,
EC TMR grant ERBFMRX-CT96-0045, INTAS grant No.96-538 and
NATO grant PST.CLG 974965. Part of this work was done while A.A.T.
was visiting the Institute of Theoretical Physics at UCSB, and he
would like to acknowledge the hospitality of ITP extended to him
and the support of the NSF grant No. PHY94-07194.

\appendix{A}{\bf The dependence of
determinants on measure and conformal factors}

\subsec{\bf Bosonic operators}

We start with a review of some general facts about
divergences and conformal anomalies of 2-d determinants.
In particular, we review the issue of the measure dependence of
the determinants
\refs{\alb, \scht}.
The action
\eqn\eqqo{\eqalign{
&S_2 = \ha \int d^2 \s \sqrt g
(g^{ij} \del_i \p \del_j \p + X \p^2)
\equiv (\p, \Delta \p)\,,
\cr
&(\p,\p') \equiv \int d^2 \s \sqrt g M(\s) \p (\s) \p' (\s)
\,,}}
where the scalar product is defined with an extra measure factor
$M$, implies that the relevant Laplace operator that occurs in
the determinant is, not $\Delta$, but rather
\eqn\meaa{
\Delta_M = M\inv ( - \nabla^2 + X). }
The dependence of $\det \Delta_M$ on $M$
can be determined \alb\ by using the standard observation that,
since $ \delta \Delta_M = - (M\inv \delta M) \Delta_M , $
 the variation of
\eqn\varofno{
\ln \det \Delta_M
= - \int^\infty_{\Lambda^{-2}} { dt \ov t}\Tr\exp(-t\Delta_M)
}
with respect to $\ln M$ can be expressed in terms of the Seeley
coefficients of $\Delta_M$. It is then easy to see that only the
quadratically (and linearly) divergent and finite parts of
$\ln \det \Delta_M$ are dependent on $M$, but that the
logarithmically divergent part is $M$-independent. This is
in agreement with naive expectation that the $M$ dependence should
be given by $\ln M$ multiplying a regularized ``$\delta(0)$''.

Equivalently, note that \meaa\ can be written as
\eqn\maa{
\Delta_M = - \td \nabla^2 + \td X,
\qquad
\td g_{ab} \equiv M g_{ab},
\qquad
\td X \equiv M\inv X
.}
Then the divergent part of this determinant is given by the
standard expression
\mac\
\eqn\staa{\eqalign{
(\ln \det \Delta_M)_\infty
=&
- {1 \ov 4\pi} \Lambda^2 \int d^2 \s \sqrt{\td g}
\pm {1 \ov 4\sqrt \pi}\Lambda \int d s \sqrt{\td \g }
\cr
&
- {1 \ov 4\pi} \ln \Lambda^2 \left[
 \int d^2 \s \sqrt{\td g} \left( {1\ov6} \td R - \td X \right)
+ {1\ov3}\int d s \sqrt{\td \g} \td K \right]\ ,
}}
where $\pm$ corresponds to the Dirichlet and Neumann boundary
conditions, $\td \g= M \g $ is the boundary metric and $K$ is the
trace of the second fundamental form. It is easy to see that the
dependence on $M$ drops out of the $\ln \Lambda$ term.\foot{This
is not unexpected since the relevant part of the logarithmic
divergence is proportional to the Euler number of the $\td g$ metric
but it should be the same as the Euler number of the $g$ metric.
Strictly speaking, this is so if $M$ is smooth
inside the domain, so as not to change the topology.
}

The dependence of the finite part of $\ln \det \Delta_M$ on $M$
is dictated by the same Seeley coefficient $a_2$ which multiples
the $\ln \Lambda$ term and determines the conformal anomaly.
This coefficient, \ $a_2$, appears in the $t\to 0$ expansion of
the heat kernel,
\eqn\diff{
\Tr[ F(\s) \exp(-t\Delta_M) ]
=
{1\ov t} a_0(F|\Delta_M)
+{1\ov \sqrt t}a_1(F|\Delta_M)
+a_2(F|\Delta_M)
+O\left(\sqrt t\right),
}
where $F$ is an arbitrary function, and
\eqn\seela{\eqalign{
a_0 (F|\Delta_M)
=&\,
{1\ov4\pi} \int d^2\s\, \sqrt g M (\s)F(\s),
\cr
a_1 (F|\Delta_M)
=&\,
\mp {1\ov8\sqrt\pi} \int d s\, \sqrt \g \sqrt{M(s)} F(s),
\cr
a_2(F|\Delta_M)
=&\,
{1\ov4\pi} \left[
\int d^2\s\, \sqrt g F(\s) b_2(\Delta_M)
+\int ds\, \sqrt \g \left( F(s) c_2(\Delta_M) \mp \ha \sqrt \g \del_n F
\right)\right]
.}}
Here
\eqn\mea{\eqalign{
b_2 (\Delta_M)
=&\, {1\ov6} \R - X - { 1 \ov 6} \nabla^2 \ln M\ ,
\cr
c_2 (\Delta_M) =&\, {1\ov3} \left( K - \ha \del_n \ln M \right)
.}}
and $\del_n$ is the inward pointing derivative normal to the
boundary. The signs $\mp$ are for D and N boundary conditions.
The dependence of the finite part on the measure factor $M$ is
found by integrating the equation (we assume that $\Delta_M$
has a trivial kernel)
\eqn\varia{
\delta ( \ln \det \Delta_M )
= - a_2(\delta \ln M | \Delta_M)
,}
i.e.
\eqn\exx{\eqalign{
(\ln \det \Delta_M)_{\rm fin}
=&\, (\ln \det \Delta_1)_{\rm fin}
-{ 1 \ov 4 \pi} \int d^2 \s \sqrt g
\left[ \ln M
\left( { 1 \ov 6} \R - X
 \right)
+ { 1 \ov 12} \del^i \ln M \del_i \ln M \right]
\cr
&\,
-{ 1 \ov 12 \pi} \int ds\,
\sqrt \g \big(\ln M \ K \mp \ha \del_n \ln M \big)
.}}

\subsec{\bf Fermionic operators}

Consider now the fermionic action
\eqn\iuy{
\int d^2\s \sqrt g\, \bar \psi e^i_\a \tau^\a D_i \psi
= \int d^2\s\sqrt g\, \bar \psi D_F \psi
,}
where $\tau_\a$ are 2-d Dirac matrices. Assume that
the norm contains an extra function ${\cal K}$
\eqn\yyyt{
\|\psi\|^2 = \int d^2 \s \sqrt g\ {\cal K}\ \bar \psi\psi
.}
Then the relevant second order operator is
\eqn\tre{
 \Delta_{\cal K}^{(F)} = ({\cal K}^{-1} D_F)^2
 = {\cal K}^{-2} ( - \hat \nabla^2 + ... )\
.}
While $\Delta_{\cal K}^{(F)}$ looks like the Laplace operator \meaa,
there are two important differences compared to the scalar case:
(i) the fermionic measure is ${\cal K}$, not $M= {\cal K}^2$, and
(ii) in addition to the overall factor ${\cal K}^{-2}$,
there is also an extra first derivative term,
leading to an extra connection and extra potential terms.

The dependence on ${\cal K}$ can be found using again the
variational argument, as in the derivation of conformal anomaly.
The logarithmic divergences again do not depend on ${\cal K}$.
The variation of $({\cal K}\inv D_F)^2 $ over ${\cal K}$ is
the same as of ${\cal K}^{-2} (D_F)^2$, but now it is the
Seeley coefficient of $({\cal K}\inv D_F)^2$ that is to be used
in \varia.

In more detail, set ${\cal K} = e^{\l}$ and choose the conformal
frame  $e^\a_i = e^\rho \d^\a_i$. Then it is easy to show that
the since the spinor derivative is
$D_j = \del_j + \ha i \tau_3 \ep_{jk} \del_k \rho$,
where the index contractions are with respect to the flat metric,
the operator $({\cal K}^{-1} D_F)^2$ becomes
(we add here a potential term $Y$ for generality)
\eqn\ffre{\eqalign{
\Delta_{\cal K}^{(F)}
=&\,
- \left(e^{-\l} e^{-{3\ov2} \rho} \tau_i \del_i e^{{1\ov2} \rho}\right)^2
+ e^{-2\l} Y
\cr
=&\,
\ - e^{-2\l-2\rho } \left(\d^{ij} + i \ep^{ij} \tau_3\right)
\del_i (1 - \l -\ha \rho)
\del_j (1 + \ha \rho)
+ e^{-2\l} Y
,}}
where $\del_i$ and $\del_j$ act on all terms to the right.
This can be put into the standard form \meaa\
(with $M={\cal K}^2$) as follows
\eqn\fre{\eqalign{
\Delta_{\cal K}^{(F)}
=&\, e^{-2\l} \left[- e^{-2\rho } (\del_k + B_k)^2 + X \right]\ ,
\cr
\del_k + B_k
=&\, \del_k + { i \ov 2} \ep_{kj} \tau_3
(\del_j \rho + \del_j \l) - \ha \del_k \l
= D_k - \ha \tau_j \tau_k \del_j \l\,,
\cr
X
=&\, Y - \ha e^{-2\rho } \del^2\rho - \ha e^{-2\rho } \del^2\l
= Y + { 1 \ov 4} \R - \ha \nabla^2 \l
\ ,}}
where we
have used that in $d=2$ \ \ $\tau^i \tau_j \tau_i =0$.

The corresponding Seeley coefficient is thus (cf. \mea)
\eqn\meal{
b_2 \left(\Delta_{\cal K}^{(F)}\right)
= \tes { \left({1 \ov 6} - {1\ov4}\right)} \R - Y
 -\tes { \left({1\ov3} - \ha\right)} \nabla^2 \l
\,,}
or, in conformal coordinates,
\eqn\mewl{
\sqrt g\,
b_2 \left(\Delta_{\cal K}^{(F)}\right)
= - \tes {\left({ 1\ov3}-{ 1\ov2}\right)} \del^2 \rho
- \left({1\ov3}-\ha\right) \del^2 \l - \sqrt g Y
,}
so that $\l$ enters like the conformal factor (this argument
determines the conformal anomaly of the Dirac operator
since $\sqrt g e^i_\a = e^{\rho}\d^i_\a$). As in \varia, we
have (omitting obvious boundary terms)
\eqn\seelvar{
\delta \left(\ln \det \Delta_{\cal K}^{(F)} \right)
= - 2 a_2\left(\delta \ln {\cal K} \big| \Delta_{\cal K}^{(F)}\right)
,}
and thus
\eqn\iexix{\eqalign{
\left(\ln \det \Delta_{\cal K}^{(F)}\right)_{\rm fin}
=&\, \left(\ln \det \Delta_1^{(F)}\right)_{\rm fin}
\cr
&\,
+{ 1 \ov 4 \pi} \int d^2 \s \sqrt g
\left[ \ln {\cal K}
\left( { 1 \ov 6} \R + 2 Y
 \right)
+ { 1 \ov 6 } g^{ij}\del_i \ln {\cal K} \del_j \ln {\cal K} \right]
.}}
Note that the conformal anomaly of a scalar is twice as of a
2-d fermion, since for a scalar $M \to e^{2\rho}$ while for
a fermion ${\cal K} \to e^{\rho}$. The anomaly of a GS spinor
is 4 times as much as of 2-d fermion since here one needs to
take ${\cal K}\to e^{2\rho}$ on top of the flat space
operator.\foot{Note that redefining spinors with careful
account of measure factors gives equivalent results. For
example, the usual 2-d spinor action is, in conformal gauge,
$\int d^2 \s \sqrt g \bar \psi e^{- { 3 \ov 2}\rho} \tau^\a
\del_\a e^{{ 1 \ov 2} \rho} \psi$
with the measure $\int d^2 \s\sqrt g \bar \psi \psi $.
Redefining $\psi' = e^{{ 1 \ov 2} \rho} \psi$ we get
$\int d^2 \s \sqrt g \bar \psi'
e^{- { 2}\rho} \tau^\a \del_\a \psi'$
with the measure $\int d^2\s \sqrt g e^{-\rho} \bar \psi' \psi'$,
i.e. ${\cal K}= e^{-\rho}$. That corresponds to the operator
${\cal K}\inv D_F = e^{- \rho} \tau^\a \del_\a $ which has the
same determinant as
$e^{- { 3 \ov 2}\rho} \tau^\a \del_\a e^{{ 1 \ov 2} \rho}$.}

It is useful to compare this with the result found by treating
$\Delta_{\cal K}^{(F)}$ as a scalar operator \mea\ with
$M= {\cal K}^2$. According to \exx\ we would get (using that
$X \to \four \R + X$ in this case)
\eqn\yexx{\eqalign{
\left(\ln \det \Delta_{{\cal K}^2}\right)_{\rm fin}
=&\, \left(\ln \det \Delta_1\right)_{\rm fin}
\cr
&\,
-{1\ov4\pi} \int d^2 \s \sqrt g
\left[ \ln {\cal K}
\left( - { 1 \ov 6} \R - 2 X \right)
+{1\ov3} g^{ij}\del_i \ln {\cal K} \del_j \ln {\cal K} \right]
.}}
Thus
\eqn\dife{
\left(\ln \det \Delta_{{\cal K}^2}\right)_{\rm fin}
- \left(\ln \det \Delta_{\cal K}^{(F)}\right)_{\rm fin}
=
-{1\ov4\pi} \int d^2 \s \sqrt g
\left[\left({1\ov3} + {1\ov6}\right)
g^{ij}\del_i \ln {\cal K} \del_j \ln {\cal K}\right]
.}
As expected, there is a non-trivial difference
for a non-constant ${\cal K}$.

\subsec{\bf Explicit results for some determinants}

As was mentioned in the text, one way to calculate the
determinants in a curved geometry is to transform
to flat metric. Instead of a complicated kinetic term
depending on induced metric one then has to deal with a
complicated mass term.

Let us consider the expression for a scalar in the general
``bent" string configuration of Section 6. The determinant
consists of two terms: a flat-space determinant
and a conformal anomaly part. For a single massless scalar
with canonical normalization the conformal anomaly related
part of its bulk effective action is
\eqn\coo{\eqalign{
\W&= \ha \ln \det (-\nabla^2) - \ha \ln \det (-\del^2)
\cr&
=- { 1 \ov 24 \pi } \bigg[ \ln \L \int d^2 \s \sqrt g \R
 +
{ 1\ov 4}
 \int \R (-\nabla^{-2}) \R \bigg]
\cr&
\to - { 1 \ov 24 \pi }\int d^2\s' \del_\a \r
\del_\a\r \ ,
}}
where we have chosen the conformal coordinate system where
\eqn\redd{
 ds^2 = e^{2\r} (d\tau^2 + d\s'^2)\,,
\qquad
\r =\ln y\,,
\qquad
d\s' = {y^2\ov y^2_0} d\s \,.
 }
To evaluate this integral let us note that at the boundary
$y=\infty$. As a result, the total derivative and boundary
contributions are trivial divergences which can be
`renormalized away' by subtracting the expression for the
straight line case ($y_0\to 0$). This allows us to freely
integrate by parts and to replace, e.g.,
$\int y^4 \to - \int y^4_0$. Dropping the total derivative
part we thus get
\eqn\conf{
\W=- { 1 \ov 24 \pi }
\int d^2\s' \del_i \r \del_i\r
 = - { 1 \ov 24 \pi }
 \int d\tau d\s \ y^{2}_0 { y'^2 \ov y^4} \ \ \to
\ \ { T L \ov 12 \pi} y_0^2
= { 2 \pi^{2}\ov 3 [\Gamma({1 \ov 4})]^{4}} { T \ov L}
\ . }
Eq. \conf\ is to be added to the flat space result \aca\
(we assume Dirichlet boundary condition)
\eqn\ffll{
 \ha \ln \det (-\del^2) = - {\pi \ov 24 }{T \ov L'}\ . }
Here $L'$ is the range of $\s'$ which is different from
$L$ by a factor,
\eqn\putr{
L' = \int d\s\ {y^2\ov y^2_0} =
 { 2 \sqrt\pi \G({5 \ov 4}) \ov \G({3 \ov 4})} {1 \ov y_0}
= { [ \G({1 \ov 4})]^2 \ov 2 \sqrt {2 \pi} } {1 \ov y_0}
= { [ \G({1 \ov 4})]^4 \ov 2 (2 \pi)^2 } L \ ,
 }
where we used that $\G({3 \ov 4}) \G({1 \ov 4}) = \sqrt 2 \pi$.
Finally, we get the following expression for the
massless scalar determinant
\eqn\gett{
\ha \ln \det (-\nabla^2) =
- { (\pi-2) \pi^2 \ov 3 [ \G({1 \ov 4})]^4 }
 {T \ov L} \ .
}
Like the flat-space potential \aca\ and like the tree-level
potential \icyc\ this expression is {\it negative}.

Multiplied by 5, this gives the result for the contribution
of massless fluctuations in the $S^5$ directions to the partition
function \parti. Other determinants should lead to similar
contributions.

Let us now consider the case of the fermionic determinant
using \iexix. One finds that the conformal anomaly for a
massless 2-d spinor is 1/2 of that for the scalar \conf, i.e.
\eqn\oonf{
\W_{\rm F}=- { 1 \ov 48 \pi }
\int d^2\s' \del_\a \r \del_\a \r
 = { \pi^{2}\ov 3 [\Gamma({1 \ov 4})]^{4}} { T \ov L}
\ . }

\appendix{B}{\bf \ \ Partition function in straight string
case ($AdS_2$) }

Here we describe a direct approach to the
 calculation of the partition function \arti\
on $AdS_2$ which complements the discussion in Section 4.3.

\subsec{\bf Spectral density and $\zeta$-function on
Poincar\'e disc}

Our starting point will be the spectrum of the operator \stsp\
with Dirichlet conditions at the boundary of the Poincar\'e disc,
following \campo.

The trace of heat kernel is defined by
\eqn\heatc{
K_B(t;m^2)
=\Tr\exp[-t(-\D+m^2)]
={1\over2\pi^2}\int_0^\inf
\exp\left[-t\lambda(\nu)\right]
\m(\nu)d\nu\ ,
}
and the zeta function is
\eqn\zetac{
\z_B(s;m^2)
=\Tr{1\over(-\D+m^2)^s}
={1\over2\pi^2}\int_0^\inf
{\m(\nu)d\nu\ov\lambda(\nu)^s}\ .
}
The density of states for a scalar Laplacian $-\D+m^2$
(in our case $m^2=0,2$)
 is
\eqn\denstb{
\mu_B(\nu)=\pi\nu\tanh(\pi\nu)\ ,
}
where the eigenvalues of the Laplacian are
$\lambda=\nu^2+m^2+\fourth.$ If we plug in this density
of states \denstb\ into \zetac\ we find (dropping the
divergence)
\eqn\tzetac{
\z_B(s;m^2)
={(m^2+\fourth)^{(1-s)}\ov 4\pi(s-1)}
-{1\over\pi}\int_0^\inf
{\nu d\nu \ov (e^{2\pi\nu}+1)\left(\nu^2+m^2+\fourth\right)^s}\ ,
}
where we have used that $\tanh(\pi\nu)=1-2/[\exp(2\pi\nu)+1]$.

The divergence in the determinant is proportional to 
$-\z_B(0;m^2)$
\eqn\zetazt{
-\zeta_B(0;m^2)
={1\over4\pi}\left(m^2+{1\over4}\right)
+{1\over48\pi}
=-{1\over4\pi}
\left({1\over 6}\R - m^2\right)\ ,
}
while the finite part
\eqn\logc{
[\ln\det(-\D+m^2)]_{\rm fin}
=-\z_B'(0;m^2)
}
is found, using \tzetac, to be
\eqn\zetpct{\eqalign{
\z_B'(0;m^2)
=&{1\over4\pi}\left(m^2+\fourth\right)
\left[\ln\left(m^2+\fourth\right)-1\right]
\cr
&+\ {1\over\pi}\int_0^\inf
{\nu d\nu \ov e^{2\pi\nu}+1}\ln\left(\nu^2+m^2+\fourth\right)\ .
}}
The coincidence limit of the propagator is (dropping
divergences)\foot{
To do the integral, we note that
${1\over e^{2\pi\nu}+1}={1\over e^{2\pi\nu}-1}-{2\over e^{4\pi\nu}-1}$,
and use $\int_0^\inf
{\nu d\nu \ov (e^{2\pi a\nu}-1)\left(\nu^2+x^2\right)}
=\ha\left[\ln(ax)-{1\over 2ax}-\psi(ax)\right]$ and
$2\psi(2x)-\psi(x)=\psi\left(x+\ha\right)+2\ln2$.
}
$$
G_B(0)=\zeta_B(-1;m^2)
=-{1\ov 4\pi}\ln\left(m^2+\fourth\right)
-{1\over\pi}\int_0^\inf
{\nu d\nu \ov (e^{2\pi\nu}+1)\left(\nu^2+m^2+\fourth\right)}
$$
\eqn\zetamo{\eqalign{=\ -\
{1\over2\pi}\psi\left(\ha+\sqrt{m^2+{1\over4}}\right)
,}}
where $\psi(x)={d\over dx}\ln\Gamma(x)$. This is equal to
$-{d\over dm^2}\z_B'\left(0;m^2\right)$.

Since for $m^2=-{1\over4}$ we can
evaluate $\zeta'(0;-\four)$ explicitly,\foot{To show this use
$\int_0^\infty {x\, dx\over e^x+1}=\ha\zeta_R(2)$,
where $\zeta_R$ is the Riemann zeta function.
and $\int_0^\infty {x\, dx\over e^x+1}\ln x
=\ha[(\psi(2)+\ln2)\zeta_R(2)+\zeta_R'(2)]$, and the identities
$\psi(2)=1-\gamma$ and $\zeta_R(2)={\pi^2\over6}$.}
we can write
\eqn\partboz{\eqalign{
\zeta'_B(0;m^2)
={1\over24\pi}(1-\gamma-\ln\pi)+{1\over4\pi^3}\zeta_R'(2)
+{1\over2\pi}
\int_{-\four}^{m^2} dx\,\psi\left(\ha+\sqrt{x+\fourth}\right)\,.
}}

The density of states of the Laplacian
 for a Majorana fermion is
\eqn\denstf{
\mu_F(\nu)=\pi\nu\coth(\pi\nu)\ ,
}
so that the $\zeta$-function is
\eqn\tzetactf{
\z_F(s; m^2)
={m^{2(1-s)}\ov 4\pi(s-1)}
+{1\over\pi}\int_0^\inf
{\nu d\nu \ov (e^{2\pi\nu}-1)\left(\nu^2+m^2\right)^s}
}
where we used that $\coth(\pi\nu)=1+2/[\exp(2\pi\nu)-1]$.
The divergence in the determinant is proportional to 
$-\z_F(0;m^2)$ and
\eqn\zetaztf{
-\zeta_F(0;m^2)
={1\over4\pi}m^2
-{1\over24\pi}
=-{1\over4\pi}\left({1\over6}\R-{1\over4}\R- m^2\right)\ .
}
This is the standard result for a 2-d fermion.
As discussed in detail in the text (and in Appendix C),
in the case of GS fermion in the conformal gauge
we should effectively multiply the $\R$ term in
$\zeta_F(0;m^2)$ by 4, ensuring the eventual cancellation
of conformal anomalies and topological infinities.

The finite part of the determinant is
\eqn\zetpctf{\eqalign{
\ln\det\left(-D_F^2\right)_{\rm fin.}
&=-\z_F'(0; m^2)
\cr
&=-{1\over4\pi}m^2\left(\ln m^2-1\right)
+ {1\over\pi}\int_0^\inf
{\nu d\nu \ov e^{2\pi\nu}-1}\ln\left(\nu^2+m^2\right)\ .
}}
The derivative of $\z$-function with respect to $m^2$
\eqn\dzetpdmtf{\eqalign{
-{d\over dm^2}\z_F'\left(0;m^2\right)
&=-{1\over 4\pi}\ln m^2
+{1\over\pi}\int_0^\inf
{\nu d\nu \ov \left(e^{2\pi\nu}-1\right)\left(\nu^2+m^2\right)}
\cr
&=-{1\over4\pi}\left({1\over |m|}+2\psi(|m|)\right)
}}
is different from the fermion propagator from \iopot\
(divided by $2m$). The expression \dzetpdmtf\ is obviously
independent of the sign of the fermion mass term.
However, supersymmetry relates fermions with opposite
masses to scalars with different masses, and the supersymmetric
regularization used in \iopot\ gave a propagator that does
depend on the sign. for the computation of the partition function,
The prescription of \iopot\ represents a different regularization
scheme and thus leads to a different expression for the
effective potential or partition function (see below).

Again, since we can calculate the finite part of the partition
function explicitly at $m=0$, we get
\eqn\zetapfriz{
\zeta_F'(0;m^2)
=-{1\over12\pi}(1-\gamma-\ln2\pi)-{1\over2\pi^3}\zeta_R'(2)
+{|m|\over 2\pi}
+{1\over 2\pi}\int_0^{m^2}dx\,\psi(\sqrt x)\,.
}

An ${\cal N}=1$ supermultiplet in $AdS_2$
contains a fermion of mass $m_F= \m$ and a boson of mass
squared $m^2_B=\m^2-\m$ \styy.
One can combine \zetaztf\ and \zetapfriz\ to get the partition function
of a single multiplet. This can be written in terms of complicated
special functions;
for $\m=\pm1$ the numerical results are
$\zeta'_B(0;0)-\zeta'_F(0;1) \sim 0.02688$ and
$\zeta'_B(0;2)-\zeta'_F(0;1) \sim 0.05269$.

\subsec{\bf ``Effective potential" in $AdS_2$ }

An alternative way to compute the partition function
is to start with the Green's functions (defined in a way
consistent with supersymmetry) and to integrate them over
the mass parameter to obtain the effective potential
as in \refs{\bul,\ina}. Following \iopot\ for a multiplet of
one boson and one fermion with masses related as in \styy\
we get
\eqn\pot{
V_{\rm eff} (\m)
= \ha \int^{m^2_B=\m^2-\m}_0 dm^2\, G(m^2)
- \ha \int^{m_F=\m}_0 dm \,2m\, G(m^2 -m)\ ,
}
where we normalized the effective potential to be zero in the case
of the massless multiplet ($\m=0$). Here
$$
G(m^2) \equiv G_B(x,x|m^2)
= \left<x\right| (-\nabla^2 + m^2)\inv \left|x\right>
$$
\eqn\stliog{= \ { 1 \ov 4\pi} \left[ - \ln(c \Lambda^2)
 + 2 \psi\left( \ha + \sqrt{ { 1 \ov 4} + m^2}\right) \right]\ ,
 }
where
$\L$ is UV cutoff. In dimensional regularization
 ($d\to 2$) we get
$\ln(c \Lambda^2) = { 2 \ov 2-d} + \ln (4\pi \L^2 a^2) - \g$.
It is assumed that fermions are treated using dimensional reduction
regularization which preserves supersymmetry, so that
the fermionic Green function satisfies
\iopot\ \
 $\tr G_F(x,x|\m) = 2 \m G_B(x,x|\m^2-\m)$.

\noindent
Then
\eqn\thennon{
-\ln Z = W_0 + \int d^2\s\ V_{\rm eff}(\m)\ ,
}
where $W_0$ is the contribution of the massless multiplet that can be
found
by integrating the conformal anomaly.
This gives an alternative way to compute the partition function.

In our case of 3 multiplets with $\m=-1$ and 5
multiplets with $\m=1$ we get a logarithmically divergent result:
the constant ($m$-independent) part of $G(m^2) $ is multiplied by
 $$
3 \left[-\ha \int^0_{2} dm^2
+ \ha \int^0_{-1} dm \, 2m \right]
+ 5 \left[ - \ha \int^0_{0} dm^2
+ \ha \int^0_{1} dm \, 2m \right]$$
\eqn\stlioin{ =\ \ha (3 -5) =-1\ .
}
The finite part is also non-zero. For a single multiplet
$$
[V_{\rm eff} (\m)]_{\rm fin}
=\,
- { 1 \ov 4\pi} \bigg[ \int^0_{\m^2-\m} dm^2\,
\psi\left( \ha + \sqrt{ { 1 \ov 4} + m^2}\right)
$$
\eqn\potr{-\
\int^0_{\m}dm\,2m\,\psi\left(\ha+\left|m-\ha\right|\right)\bigg]
=\,
{1\ov4\pi}\int^0_{\m}dm\,\psi\left(\ha+\left|m-\ha\right|\right)
,}
where we have changed the integration variable
in the first term ($m^2 \to m^2 -m$).\foot{The observation
that for a supermultiplet the two terms combine in this
way was made in \iopot\ and implicitly in \barf.
Notice that since $\int^a_b dm\ \psi(m) = \ln \G(a) - \ln \G(b),$
the resulting integral is easily computable
by splitting the interval and changing the variables.}
For $\m=-1$ we get explicitly
\eqn\potor{
[V_{\rm eff} (-1)]_{\rm fin}
={1 \ov 4\pi} \int^0_{-1} dm\, \psi( 1-m)
={1 \ov 4\pi} \int^2_{1} dm\, \psi( m) =0
.}
For $\m=1$ we get zero bosonic contribution and thus
\eqn\potr{\eqalign{
[V_{\rm eff} (1)]_{\rm fin}
=&\,
- {1\ov4\pi}\int^1_{0} dm\,2m\psi
\left(\ha+\left|m-\ha\right|\right)
\cr
=&\,
- { 1 \ov 4\pi} \int^1_{1/2} dm\, 2 \psi( m)
= { 1 \ov 4\pi} \ln \pi
,}}
 where we split the integral into two parts and
 changed the variable.

For 3+5 multiplets we get a non-zero result (note that the
first term is actually zero)
\eqn\fgy{
3 [V_{\rm eff} (-1)]_{\rm fin} + 5 [V_{\rm eff} (1)]_{\rm fin}
={ 5 \ov 4\pi} \ln \pi\ .}

The contribution of the 8 massless multiplets can be found
in the conformal frame $ds^2 = w^{-2} (dt^2 + dw^2)$ to be
\eqn\vool{\eqalign{
W_{0}
=&\,
- 8 \times\left(1+\ha\right)\times{1\ov24\pi}
\int d^2\s \sqrt g g^{ij}\del_i \r \del_j\r
\cr
=&\,
-{1\ov2\pi}\int d^2\s\sqrt g\
,}}
where $g^{ww}\del_w \r \del_w \r=1, \ \r= - \ln w$.

\appendix{C}{\bf \ \ Comments on 2-d determinants}

In the main text we have performed (following earlier discussions
of the GS string in \refs{\polu,\sedr,\kara,\leu,\wig,\lech})
a local rotation to put the GS action \frmo\ in a ``2-d fermion
in curved 2-d space'' form. In general, the resulting Jacobian
is non-trivial and is given by a Polyakov-Wiegmann \pow\
expression which is a $U(1)$ WZ action.

In the case when 2-d metric is kept independent the account of
the contribution of the rotation Jacobian is crucial in order
to show that the conformal anomaly of a GS fermion is 4 times
the naive anomaly of a 2-d fermion -- as needed to cancel the
conformal anomaly of GS string in flat space. As was already
mentioned in the text, the cancellation of (dilatonic part of) the
conformal anomaly in the one-loop approximation we considered
is exactly the same as in flat space (the curved space-time background
changes the $O(\R)$ conformal anomaly only starting with
the 2-loop approximation \FT).

Let us summarize some basic facts about these determinants.
They are always defined modulo local counterterms of background
fields which are to be chosen consistent with the symmetries
that are to be preserved (see \leu). For generality, we consider
the chiral case, the non-chiral one is just a combination of the
two chiral ones. The standard 2-d Weyl spinor operator is (we use
the Euclidean notation where $\tau_{1,2} = \s_{1,2}$ are the Pauli
matrices; here $k,n,m=1,2$)
\eqn\compa{
D(e) = \ha i ( 1 -\tau_3) e^k_\a \tau^\a ( \del_k + i \tau_3 w_k)\,,
\qquad
\om^{\a\b}_k = 2 \ep^{\a\b} w_k\,.
}
Then
\eqn\compb{
\ln \det D(e) = - { i \ov 12 \pi} I( i w)
= { 1 \ov 2 \times 96 \pi} \int
 \R \left( - \nabla^2\right)\inv ( \R - 4 i \nabla_k w^k)
\,.}
In the case of an abelian gauge field background in flat space
\eqn\compc{
I(A) = W(T_A, A) = \int d T_A T_A\inv \wedge A \,,
\qquad
(\sqrt g g^{mn} + i \ep^{mn}) (\del_n + A_n ) T_A=0 \,.
}
In conformal coordinates
$A_z = - \del_z \l$,\ $\l= \ln T_A$, so that
$I(A) = \int \del_z \l (\bar \del_z \l - A_{\bar z}), $
which
 is different from usual abelian expression $\int \del \l \bar \del
\l$
by the local counterterm $\int A_z A_{\bar z}$.\foot{An equivalent
(up to a local counterterm) expression is
$$
\ln \det D = - { 1 \ov 3} Z(w) + Z(A) \,,
\ \ \ \
Z(B) =
- {1\ov 4\pi} \int \ep^{mn} \del_m B_n \left(-\nabla^2\right) \inv
[ \ep^{mn} \del_m B_n - i \nabla_m B^m ]
\, . $$
Using the
conformal gauge and definitions
$
g_{mn} = e^{2 \r} \d_{mn} \,,
\
w_m = \del_m \l + \ha \ep_{mn} \del^n \r \,,
\
A_m = \del_m a + \ep_{mn} \del^n b \,,
$
we get
$$
\ln \det D (e,A) = { 1 \ov 4\pi} \int d^2 \s \left[
{1\ov12} \del^m \r \left( \del_m \r + \ha i \del _m \l\right)
 - \del^m b \left( \del_m b + \ha i\del _m a\right)\right]
\,.
$$
}
The gauge field dependent WZ term in the Dirac operator case
is proportional to $I(A) + I(\bar A)$, or
\eqn\compd{
i \int \left( A {\bar \del \ov \del} A
+ \bar A { \del \ov \bar\del} \bar A \right)\ ,
}
or, after adding a local $A \bar A$ term,
$- i\int (\del \bar A - \bar \del A)
 { 1 \ov \del \bar \del} (\del \bar A
-\bar \del A)$.

For a Majorana spinor on a curved background only the first
(Polyakov) conformal anomaly term is present in \compb\
(with coefficient which is 1/2 of the scalar one), while the
imaginary Lorentz-anomaly term cancels out. The gravitational
WZW action $I$ is simply $\sim \int \del^k b \del_k b$,
where $w^k = i\ep^{kn} \del_n b $, up to a local counterterm
$\int w^k w_k$.

If there is also some internal connection acting on flavor
indices of fermions, then under the chiral projector in 2
dimensions it can be formally rotated away (this is clear
in conformal coordinates)
\eqn\compe{
D(e,A) = \half
 i ( 1 -\tau_3) e^m_\a \tau^\a ( \del_m + i \tau_3 w_m + A_m )
 = T_A D(e) T\inv_A \,,
}
where $T_A$ is a local rotation ``eliminating" $A_m$, and
\eqn\compf{
\ln \det D(e,A) = - { i N\ov 12 \pi} I( i w) + { i \ov 4 \pi} I( A)
\,.}
Now, consider a more general case
we are actually interested in
\eqn\compg{
D(\r,B) = \ha i (1 - \r_3) \r^m (\del_m + B_m)
\,,}
where
$\r_m = \r_m (x) $ is an arbitrary $2N \times 2N$
representation of the 2-d Dirac algebra for some metric $g_{mn}(x)$,
 satisfying
\eqn\comph{
\r_{m} \r_{n} = g_{mn} + i e_{mn} \r_3\,,
\qquad
e^{mn} = { 1 \ov \sqrt g} \ep^{mn}
\,.}
The condition on the connection $B_m$ is
\eqn\compi{
\del_m \r_n + [B_m, \r_n] - \G^k_{mn} \r_k = 0
\,.}
Since $g_{mn} = e^\a_m e^\b_n \d_{\a\b}$, there exists a local
rotation $S$ that transforms $\r_m$ into the constant
$2 \times 2$ Pauli matrices times $N \times N$ unit matrix times
the zweibein $e^\a_n$
\eqn\compj{
S \r_n S\inv = \tau_\a \times I\ e^\a_n
\,.}
Then
\eqn\compk{\eqalign{
S D(\r, B)S\inv
=&\, D(e, A) \,,
\cr
\del_m + i \tau_3 w_m + A_m
=&\, S (\del_m + B_m) S\inv
\,, }}
and hence \leu\
\eqn\then{
\ln \det D(\r,B) = - { i N\ov 3 \pi} I( i w)
 + { i \ov 8 \pi} I( B)
\,. }
Note that for $B_m= i w_m$ (with extra 2 flavors, i.e.
$I( B) = 2I( i w)$) and $N=1$ we get back to \compb, i.e.
$ - { i \ov 12 \pi} I( i w)$.

For the flat target space \sedr\ ($B_m$ comes from integration
by parts as required for self-adjointness of the Dirac operator)
\eqn\compl{
\r_n =\del_n x^a \G_a \,,
\qquad
B_m = \ha \r_3 \del_m \r_3 = - \ha \r_n \nabla_m \r^n
\,.}
In the case of the non-chiral Dirac fermions (where there is no
problem with a definition of the determinant of $D(\r,B)$
discussed in \leu) we get simply
\eqn\compm{
\ln \det D(\r, B )
=
 { N \ov 24 \pi} \int
 \R \left( - \nabla^2\right)\inv \R
+ { i \ov 8 \pi} [ I( B) - \td I (B) ]
\,,}
where $ \td I $ is defined by \compc\ with parity-reflected
condition. In the abelian case $I( B) - \td I (B) =0$,
up to a local counterterm. Notice that the conformal anomaly term
(here we discuss the Dirac spinor, so that it is twice that of
a Majorana GS fermion) is 4 times bigger than for the usual 2-d
Dirac fermion.

\appendix{D}{\bf Superstring partition function
 in $AdS_3\times S^3$ with RR flux}

The same calculation can be done in the case of a superstring
in the $AdS_3 \times S^3 \times T^4$ with RR background. The
corresponding GS action has the form very similar to \adss\
one and was discussed in \adtr. Since all the classical solutions
discussed above depend on only three of the $AdS_5$ coordinates,
they can be directly embedded in $AdS_3$ and are still minimal
surfaces.

There are some differences in treating the quadratic fluctuations
in $AdS_3\times S^3\times T^4$:

(1) there are no massive fluctuations in the $x^2,x^3$
directions which are replaced by extra two massless
fluctuations in the toroidal $T^4$ directions.

(2) only half of the 8 effective 2-d fermions get $\s_3$ mass
term in \fuum, \lferco\ and \fuumqaq. In the present case it is
natural to split the $\G$-matrices in $(3+3)+4=6+ 4 $ way,
$\G_{a } =\g_a \times I_4$,\ $a=0,1,...,5$, where $\g_a$ are
6-d $8\times 8$ matrices. The `mass term' in the covariant
derivative \cov\ and in the string action now originates from
the sum of the electric and magnetic RR 3-form field strengths,
$\del x^\m \del x^\n \bar \theta \G_\n \G^{abc} \G_\m
(F_{abc} + F^*_{abc})$.
The combination of the field strengths produces the $(1+\G_7)$
projection operator, so that only half of the 6-d fermions get a
mass term.

For example, in the case of the straight line the minimal surface
is still $AdS_2$, but the set of fluctuation fields is different.
There are 7 massless bosons, one mass 2 boson, four massless
fermions and four mass 1 fermions. They form one $\cN=4$ multiplet
(the dimensional reduction of the $\cN=2$ vector multiplet in
$D=4$) with three massless and one massive boson and four massive
fermions. There are 4 other massless bosons and four massless
fermions which can also combine into $\cN=4$ multiplet. One can
also check that the vacuum energy as defined by the zeta function
again vanishes for this combination of fields. The same conformal
anomaly calculation as in section 3.4 gives that $Z=1$.

In the case of the general bent string configuration
(parallel Wilson lines), the analog of the \adss\ partition
function \parti\ takes the form (in static gauge)
\eqn\parti{
Z_{AdS_3 \times S^3 \times T^4}^\parallel
={
\det^{4/2}\left( - \hat \nabla^2 + \four \R + 1\right)
\det^{4/2}\left( - \hat \nabla^2 + \four \R \right)
\over
\det^{1/2}\left( - \nabla^2 + \R + 4\right)
\det^{7/2}\left( - \nabla^2 \right)}
\,.}

\listrefs
\bye